\PassOptionsToPackage{table}{xcolor} 
 \documentclass[manuscript]{acmart}

\usepackage{subcaption}
\usepackage{pifont}

\usepackage{tcolorbox}
\newcommand{\fixme}[1]{{\color{black} #1}}
\usepackage{todonotes}

\AtBeginDocument{%
  }

\begin{document}

%%\title[Generative AI for Accessibility Design of Shopping Websites]{From Cluttered to Clear: Improving the Web Accessibility Design for \\Screen Reader Users With Generative AI}
\title[Generative AI for Accessible Design of Shopping Websites]{LLM-Driven Optimization of HTML Structure to Support Screen Reader Navigation}
%% Improving Web Accessibility for Screen Reader Users in E-commerce with Generative AI.

\author{Yaman Yu}
\email{yamanyu2@illinois.edu}
\affiliation{%
  \institution{University of Illinois at Urbana-Champaign}
  % \streetaddress{P.O. Box 1212}
  \city{Champaign}
  \country{USA}
}
\author{Bektur Ryskeldiev}
\affiliation{%
  \institution{Mercari R4D}
  \city{Tokyo}
  \country{Japan}
}

\author{Ayaka Tsutsui}
\affiliation{%
  \institution{University of Tsukuba}
  \city{Tsukuba}
  \country{Japan}}

\author{Matthew Gillingham}
\affiliation{%
  \institution{Mercari R4D}
  \city{Tokyo}
  \country{Japan}
}

\author{Yang Wang}
\email{yvw@illinois.edu}
\affiliation{%
  \institution{University of Illinois at Urbana-Champaign}
  % \streetaddress{P.O. Box 1212}
  \city{Champaign}
  \country{USA}
}
%% who coauthors

\renewcommand{\shortauthors}{Anon et al.}

\begin{abstract}
Blind and low vision users often face significant barriers when navigating online shopping websites using screen readers. Complex layouts, unclear content hierarchies, and visually driven designs create a frustrating and inefficient browsing experience, particularly on unfamiliar platforms. While prior accessibility tools focus on isolated elements such as product descriptions or image alt text, they often fall short of addressing the structural and navigational challenges screen reader users encounter across entire webpages. In this work, we explore how Generative AI (GenAI) can be leveraged to improve the accessibility of shopping websites by automatically restructuring their HTML content. We conducted a three-phase study: formative interviews with screen reader users, system development of a GenAI-powered browser extension, and user evaluation through both automated audits and real-world testing. Our tool dynamically reorganizes web content to better align with screen reader navigation patterns. Results from user studies with blind and low vision participants show that the GenAI-generated pages significantly improve navigation efficiency, content clarity, and overall usability. Participants highlighted benefits such as more logical section order and reduced browsing fatigue. Our findings demonstrate the potential of GenAI to support comprehensive, user-centered accessibility improvements directly within the structure of existing websites.

\end{abstract}



%\keywords{accessibility, artificial intelligence, generative AI, screen readers, web accessibility, e-commerce}



\begin{teaserfigure}
    \includegraphics[width=\textwidth]{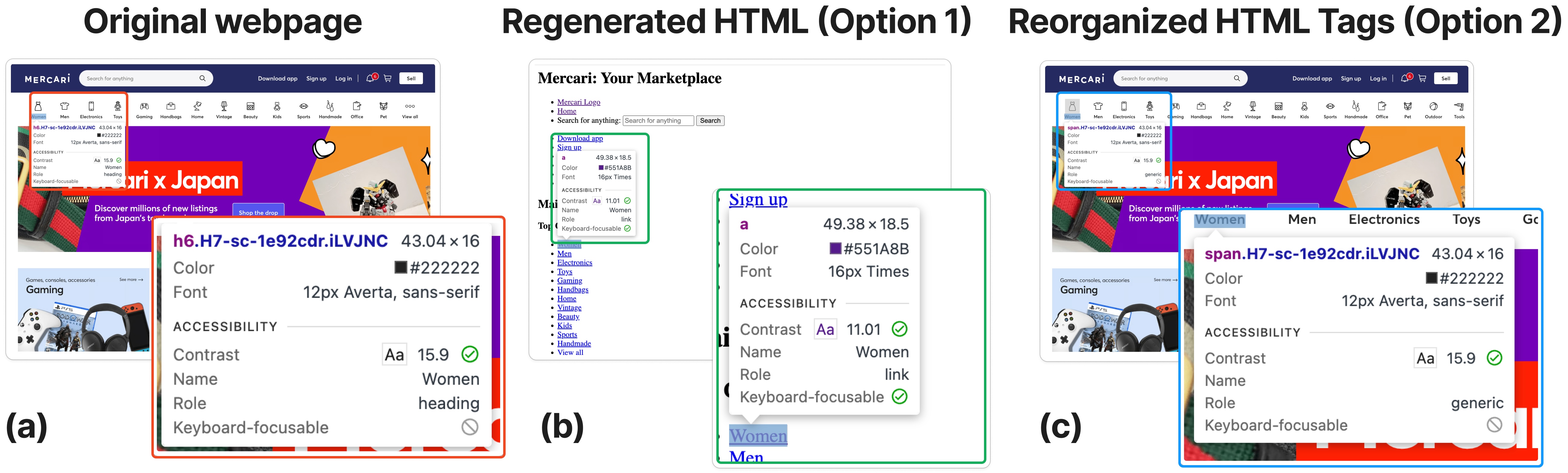}
    \caption[This image compares three versions of a web page: \textbf{(a)} the original website, \textbf{(b)} Option 1 (Regenerated HTML), and \textbf{(c)} Option 2 (Reorganized HTML Tags). The original website, shown on the left, is the Mercari marketplace featuring a colorful layout with product categories, a search bar, and a prominent ``Mercari x Japan'' banner. In the center, Option 1 displays a text-only version of the site, with a hierarchical structure of headings and links, removing visual elements to focus on content accessibility. On the right, Option 2 appears visually identical to the original but with potential structural changes to improve screen reader navigation. Each version includes a zoomed-in section highlighting accessibility information: the original shows a ``Women'' heading with contrast ratio and role details; Option 1's zoomed section displays a ``Sign up'' link with improved contrast and keyboard accessibility; Option 2's zoomed area is similar to the original but with a ``generic'' role instead of ``heading''. This comparison illustrates different approaches to enhancing web accessibility: complete HTML regeneration for a text-focused experience, and HTML tag reorganization for improved structure while maintaining visual design.]{\fixme{From left to right: \textbf{(a)} the original website, \textbf{(b)}  Option 1 (Regenerated HTML), and \textbf{(c)}  Option 2 (Reorganized HTML Tags). Option 1 is generated using a GenAI-powered extension that rewrites the HTML to enhance accessibility, while Option 2 reorganizes the original HTML tags to address accessibility issues without altering the visual design. This example highlights the changes made in both versions compared to the original website. Both versions simplify navigation by removing single-category headings, reducing clutter and minimizing fatigue for screen reader users.}}
    % \ {briefly say here how option 1 and 2 are generated}}}
    \label{fig:teaser}
\end{teaserfigure}

\maketitle

\section{Introduction}

People who are blind or have low vision rely on the Internet to access information and perform everyday tasks, such as shopping for essential items and managing personal finances \cite{10.1145/3234695.3236337}. However, many blind or low vision users face significant challenges when navigating websites with screen readers, which are tools that convert text to speech for auditory access \cite{10.1145/2207676.2207736}. Common issues include inaccessible table formats \cite{10.1145/274497.274521}, inadequate image descriptions that lack important visual context \cite{10.1145/3173574.3174092}, insufficient navigation aids \cite{10.1145/3173574.3173585}, and confusing or unclear explanations \cite{8651676, 10.1145/3167132.3167349}. These accessibility barriers can lead to frustration and a sense of helplessness for blind or low vision users when interacting with websites \cite{baker2005building}.

For blind or low vision users, navigating unfamiliar websites can be particularly challenging due to the need to understand and interact with unique website elements and layouts, which can be confusing and time-consuming. This problem is especially evident on shopping websites, where visually appealing but complex designs often take precedence over accessibility \cite{haubl2000consumer}. Shopping websites frequently use intricate layouts, dynamic content, and numerous filters or categories that may not be accessible to screen readers, further complicating the navigation experience for blind or low vision users. In addition, shopping tasks such as finding product details, accessing reviews, managing shopping carts, and checking out involve multiple steps and specific interactions that are difficult to execute without clear and consistent navigation aids. Although simplified website designs are recommended to improve accessibility \cite{10.1145/1978942.1979001}, the lack of inclusive design on many online shopping platforms makes it difficult for blind or low vision users to perform common tasks, such as comparing products across multiple sites, exploring deals or reading detailed product information.

Researchers have explored various methods to improve web accessibility for blind or low vision users within the field of Human-Computer Interaction (HCI). Studies have focused on creating interactive systems to help users navigate specific elements on product pages or retrieve essential information using AI-based assistants~\cite{10.1145/3411764.3445547, 10.1145/3234695.3236337}. However, these existing solutions primarily target specific elements, such as product descriptions or image retrieval, rather than ensuring a comprehensive, end-to-end accessible experience across the entire web page, particularly for users who may be unfamiliar with a new platform. They often rely on external tools or assistants that interact with existing site structures, which may not fully align with users' browsing needs and preferences. To fill this gap, we propose using Generative AI (GenAI) to automate the optimization of shopping websites for screen reader users. Our research goes beyond previous efforts by focusing on improving the overall structure and content of websites, ensuring a more inclusive and user-friendly experience for blind or low vision users using a screen reader. Our study aims to answer the following research questions (RQs):
\begin{enumerate}
    \item[\textbf{RQ1:}] What specific accessibility barriers do screen reader users face on shopping websites?
    \item[\textbf{RQ2:}] How can GenAI be utilized to improve the accessibility of these websites?
    \item[\textbf{RQ3:}] How effective is GenAI-based solution in enhancing accessibility?
\end{enumerate}

To address these research questions, we conducted a three-phase study that involved formative interviews, system development, and user evaluations. In the formative study, we interviewed six screen reader users to identify the specific challenges they face on shopping websites and to understand their current strategies to overcome these barriers. The insights gained from this phase informed the development of a GenAI-based tool that automatically revises HTML content to improve accessibility for screen reader users. The tool was implemented as a browser extension that dynamically reorganizes web page elements to align with screen reader navigation patterns. Finally, we evaluated the tool through automated accessibility audits and user testing on Mercari, a legitimate and unfamiliar English-language shopping site selected to minimize bias from participants' prior experience.

Our findings indicate that the proposed GenAI solution significantly improves accessibility by restructuring web page content in ways that make it more intuitive and easier to navigate for blind or low vision users. In user evaluations, seven screen reader users tested both the original and revised versions of a shopping website. The results demonstrated that our tool's regenerated HTML version led to more efficient navigation and a clearer understanding of website hierarchy compared to the original version. The participants particularly appreciated the logical reordering of sections, the addition of summary headings, and enhanced navigation flexibility, highlighting the potential of our approach to provide a more comprehensive and inclusive web experience for blind or low vision users.

Our study makes several important contributions to the field of web accessibility. First, we provide an in-depth analysis of the specific accessibility barriers that screen reader users encounter on shopping websites, particularly on non-mainstream platforms with complex layouts and dynamic content. This analysis extends the current understanding of the unique challenges faced by blind or low vision users in navigating shopping environments, such as inconsistent header navigation across different websites, browsing and item comparison fatigue, and how blind or low vision users adopt the latest GenAI technologies to complement the lack of product information in images and description. Second, we introduce the application of GenAI for enhancing web accessibility. Unlike previous approaches that focus on isolated elements or rely on external tools, our GenAI-based solution offers an automated, comprehensive optimization of website structure and content directly on the web page. Third, we developed a browser extension and validated that GenAI can improve the accessibility of shopping websites through evaluations using both automated tools and screen reader users. Finally, our user evaluations show that the revised HTML content generated by our extension leads to substantial improvements in usability and navigation efficiency for screen reader users.
\section{Related Work}
% This section clarifies the challenges faced by blind or low vision users who utilize screen readers for navigating websites and distinguishes our research from previous HCI studies. Section \ref{subsection_2.1} identifies the difficulties encountered when using unfamiliar websites, with a particular focus on the accessibility of shopping sites that feature complex web designs. Section \ref{subsection_2.2} reviews existing systems for improving the accessibility of online shopping platforms, particularly those using command-based interaction or restructured product pages. Section \ref{subsection_2.3} outlines broader AI-assisted tools that enhance existing web content, including image descriptions and accessible tables, and highlights their limitations for new site development. Section \ref{subsecction_2.4} discusses approaches for web accessibility and design using AI and compares these with our research.

\subsection{Online (Shopping) Accessibility Challenges}
\label{subsection_2.1}
Efforts to improve shopping accessibility can be categorized as offline and online. Offline challenges have been addressed through third-party assistance \cite{offlineshopping1} and integrated technologies \cite{offlineshopping2,offlineshopping3}, yet blind or low vision users with mobility limitations still face difficulties visiting physical stores independently \cite{10.1145/2971648.2971723, 10.1145/2513383.2513449}. As a result, many rely on websites for everyday shopping and information gathering \cite{10.1145/3234695.3236337}. However, online shopping presents its own barriers. Navigating unfamiliar websites is often time-consuming and frustrating for screen reader users due to visual-centric design and mouse-oriented interfaces \cite{10.1145/1805986.1806005, 10.1145/2207676.2207736}. Screen readers may misread or skip important content, creating a sense of confusion and helplessness \cite{baker2005building}. Prior studies have identified key accessibility issues such as missing navigation aids \cite{10.1145/3173574.3173585}, poorly designed tables \cite{10.1145/274497.274521}, lack of alternative text \cite{10.1145/3173574.3174092, 10.1145/2470654.2481291, 10.1145/1866029.1866080, 10.1145/2702123.2702437}, and overly verbose descriptions \cite{8651676, 10.1145/3167132.3167349}. These problems limit access to online information resources \cite{leuthold2008beyond, 10.5555/1162223}. Design inconsistencies across websites further increase the burden, particularly for new users or those exploring unfamiliar platforms \cite{10.1145/1978942.1979001}. Shopping websites, with their complex layouts and varied content types \cite{haubl2000consumer}, intensify these challenges. Inadequate image descriptions and cluttered structures often make it difficult to retrieve information effectively \cite{10.1145/3313831.3376404, webaim_survey}. To address these issues, HCI researchers have explored various strategies to improve web accessibility for visually impaired users, as detailed in the following sections.

\subsection{\textbf{Improving Web Browsing Accessibility}}
\label{subsection_2.2}
Prior work has introduced tools to improve web accessibility for visually impaired users, particularly in online shopping contexts. For instance, Wang et al. developed a browser extension for Amazon that extracts product descriptions to help visually impaired users understand item appearance through a question-answering system~\cite{10.1145/3411764.3445547}, while Stangl et al. created a system that allows screen reader users to ask questions about products and receive specific, relevant information~\cite{10.1145/3234695.3236337}. These systems highlight the benefits of command-based interaction and targeted information retrieval.  However, they often introduce a steep learning curve, requiring users to adapt to an additional interface layer instead of browsing the website naturally. Concerns about trust and the reliability of extracted information also remain. Beyond shopping-specific tools, broader web accessibility efforts have focused on enhancing existing content. WebInSight \cite{bigham2006webinsight} and SADIe \cite{harper2007sadie} insert alternative text or semantic markup, while other tools address specific components, such as command-based browsing \cite{10.1145/3519032} and table conversion \cite{10.1145/3491102.3517469}. Although useful, these approaches address isolated accessibility issues and do not improve the overall content structure. Since effective screen reader navigation depends on coherent page-wide hierarchy and logical content flow, isolated fixes often fall short of supporting seamless interaction. 

\subsection{AI-Driven Web Accessibility and Design}
\label{subsecction_2.4}

Recent work has explored AI-driven methods to improve web accessibility, including restructuring data tables \cite{10.1145/3491102.3517469}, converting PDFs to HTML \cite{10.1145/3441852.3476545, wang2021improving}, and generating image and audio descriptions \cite{huh2023genassist, lei2020mart, 10.1145/3411764.3445347, 10.1145/3357236.3395433, 10.1145/3334480.3382821}. Tools have also supported video comprehension for visually impaired users \cite{van2024making}. Due to the time-intensive nature of manual remediation \cite{9809423}, automated approaches—such as ChatGPT-based accessibility repair \cite{10.1145/3594806.3596542}, evaluation tools \cite{10.1145/3650212.3652113}, and context-aware image description systems \cite{10.1145/3663548.3675658}—have gained traction. However, LLMs face challenges such as hallucinations \cite{bigham2017effects, 10.1145/3663548.3675631} and the risk of introducing new barriers when not designed with users in mind, as seen in accessibility overlays that disrupt screen readers \cite{10.1007/978-3-031-08645-8_2, 10.1145/3663548.3675650}. Recent studies have begun addressing accessibility in specific contexts like C2C shopping, improving communication and listing quality \cite{10.1145/3604571.3604586, 10.1145/3517428.3550390}.

Building on this direction, our research proposes a GenAI-based approach to automate accessibility design for shopping websites. Unlike prior efforts focused on isolated HTML fixes \cite{10.1145/1866029.1866041, 10.1145/1414471.1414508}, we aim to use LLM to regenerate or reorganize the HTML of existing websites, creating a clearer information hierarchy and facilitating navigation that aligns with screen reader users' natural browsing behaviors, without introducing an additional layer to learn.

\section{Methodology Overview}

\begin{figure}[h]
\centering
\includegraphics[width=13cm]{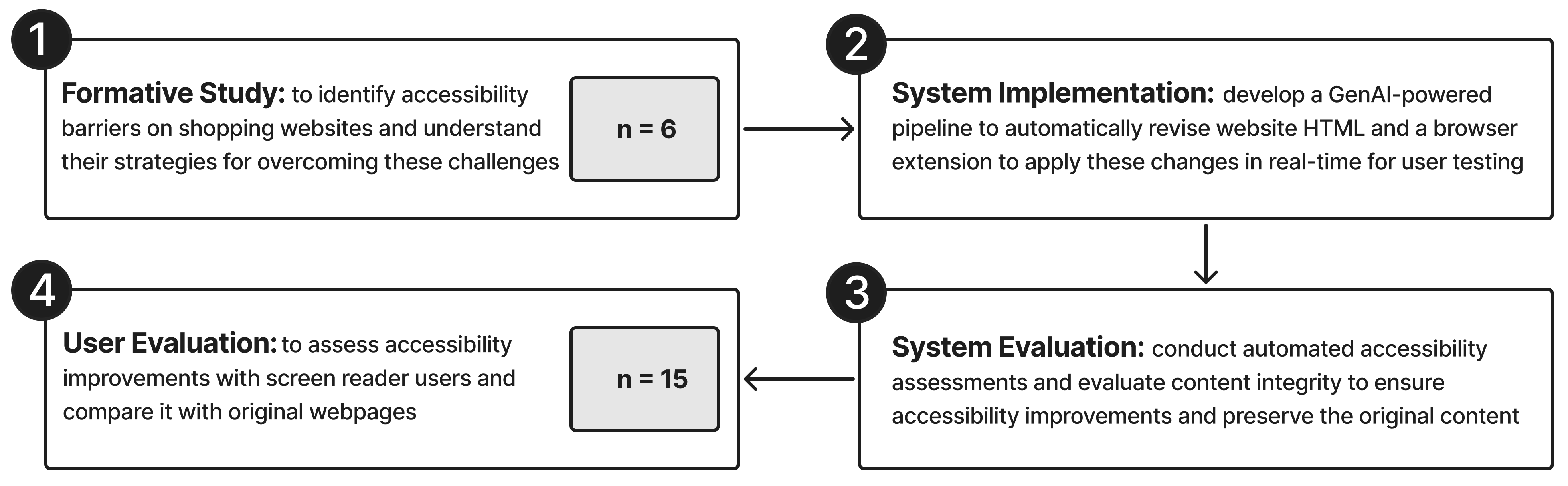}
\caption[A flowchart depicting the four main steps of the study methodology. Step 1: Formative study to identify challenges, coping strategies, and needs of screen reader users navigating shopping websites. Step 2: Development of a browser extension system powered by GenAI to automatically revise website HTML for improved accessibility. Step 3: Technical assessments to evaluate accessibility improvements and content integrity of the revised webpages. Step 4: User evaluations to validate accessibility enhancements and gather qualitative feedback. Arrows connect the steps sequentially, showing the flow from formative research to development, technical assessment, and user validation.]{
\textbf{\fixme{Methodology Overview}---%
    \textmd{{\small Note that participants of the user evaluation (step \ding{185}) do not overlap with those in steps \ding{182}}. 
    }
 }
 }
\label{fig:method}
\vspace{-0.16in}
\end{figure}

\fixme{To address our research questions (RQ), we designed a structured methodology comprising four key steps, summarized in Figure~\ref{fig:method}. In step \ding{182}, we conducted a formative study to uncover the challenges screen reader users face, their coping strategies, and their specific needs when navigating shopping websites. These findings informed step \ding{183}, where we developed a browser extension system powered by GenAI to automatically revise website HTML for improved accessibility. In steps \ding{184} and \ding{185}, we evaluated the improved webpages through technical assessments and user evaluations to validate both their accessibility enhancements and content integrity.}
%We conducted a three-phase study to address the proposed research questions:

% recover

\paragraph{\textbf{Participants recruitment.}}
\label{sec:recruit}
A total of \fixme{21} participants from the United States participated in our study, with 6 in the formative study and \fixme{15} in the user evaluation. Participants were recruited via a screening survey shared on platforms like Prolific\footnote{\url{https://www.prolific.com/}}, as well as mailing lists and word of mouth. The survey assessed vision ability, screen reader usage, and online shopping frequency to identify blind or low vision users of shopping websites. User evaluation participants were distinct from those in the formative study to avoid priming bias. Each participant received a \$25 Amazon gift card per hour. Study sessions were audio-recorded, transcribed, and conducted with participants’ consent. The study was approved by the Institutional Review Board (IRB).

\section{Formative Study}
\label{sec:phase1}
\label{sec:formative study}
Prior research has examined the accessibility barriers that blind or low vision users face on commonly used shopping websites~\cite{10.1145/3313831.3376404, 10.1145/3411764.3445547}, primarily focusing on identifying challenges and improving image accessibility. One study~\cite{10.1145/3234695.3236337} specifically noted that blind or low vision users experience difficulties accessing product information due to missing alt text for images and non-compliance with WCAG standards \cite{wcag21}. However, there is limited research on the specific challenges that blind or low vision users face throughout their shopping process due to website design and development, especially on new or unfamiliar websites. Addressing this gap is crucial, as blind or low vision users often want to explore a broader range of shopping options and compare products but are limited to a very narrow selection of accessible websites. There is also a need to understand their strategies for overcoming these challenges and their desired solutions to address these issues. We conducted a formative study to explore these questions.

% Based on the observed literature, we outline two primary research questions: (1) ``\textbf{What problems do screen reader users experience when shopping online?}'' and (2) ``\textbf{How can generative AI applications address these problems?}'' 

% \begin{enumerate}
%     \item \textbf{Initial investigation into challenges experienced by screen reader users} when browsing online shopping websites,
%     \item \textbf{Developing a browser plugin prototype} that analyzes a website and prompts a Large Language Model (LLM) to restructure or rewrite a website to make it more suitable for screen readers,
%     \item \textbf{Prototype validation} in a series of user tests followed by semi-structured interviews
% \end{enumerate}

% \subsection{Initial Investigation Into Screen Reader Issues}

% \subsubsection{Study Design}
%introduction
\subsection{Participants Demographics}
The formative study included six blind or low vision users who regularly used screen readers for online shopping. These participants, aged 25 to 54 years, were recruited through social networks and mailing lists. Half of them were legally blind (n = 3), and the other half had significant visual impairments requiring assistive technologies (n = 3). Four participants self-identified as male and two as female. Beyond basic demographics, we collected detailed background information on their online shopping habits to ensure a diverse group of participants. This included people who shop online once a week and those who did so 2-3 times a week, on various e-commerce platforms. More details are provided in the demographic Table \ref{tab:phase1demo}.

% recover
\begin{table}[H]
\begin{tabular}{llllll}
\hline
\textbf{ID} &
  \textbf{Age} &
  \textbf{Gender} &
  \textbf{Visual Ability} &
  \textbf{\begin{tabular}[c]{@{}l@{}}Online shopping\\ Frequency\end{tabular}} &
  \textbf{\begin{tabular}[c]{@{}l@{}}Used Screen\\ Reader\end{tabular}}  \\ \hline
\textbf{P1} & 25-34 & Male   & Completely blind           & Once a week      & NVDA, Google Talkback                \\
\textbf{P2} & 25-34 & Male   & Blind                      & Once a week      & JAWS, NVDA, VoiceOver \\
\textbf{P3} & 35-44 & Male   & Totally blind              & Once a week      & JAWS, VoiceOver                \\
\textbf{P4} & 25-34 & Female & I have light perception    & Once a week      & JAWS, NVDA, VoiceOver                \\
\textbf{P5} & 25-34 & Male   & Only some light perception & 2-3 times a week & JAWS, NVDA, VoiceOver                 \\
\textbf{P6} & 45-54 & Female & Limited light perception   & Once a week      & JAWS, NVDA           \\ \hline
\end{tabular}
\caption[Table 1. Formative Study Participants Demographic Table. The table contains demographic information for six participants labeled P1 through P6. It has six columns: ID, Age, Gender, Visual Ability, Online shopping Frequency, and Used Screen Reader. The participants’ ages range from 25-54, with four males and two females. Their visual abilities vary from completely blind to limited light perception. All participants shop online at least once a week, with one shopping 2-3 times a week. They use various screen readers, including NVDA, JAWS, Google Talkback, and VoiceOver, with most participants using multiple screen readers. This table provides a comprehensive overview of the diverse group of visually impaired participants in the formative study, highlighting their varied demographics and screen reader usage patterns.]{Formative Study Participants Demographic Table}
\label{tab:phase1demo}
\end{table}

\subsection{Semi-Structured Interview Procedure} 
The interviews were conducted online using the Zoom video calling platform and lasted between 45-60 minutes. At the beginning of each interview, we obtained participants' consent for the study and for recording the session. The interview was divided into two parts. In the first part, we asked participants about their experiences with online shopping, the challenges they encountered, the accessibility issues they faced on previous shopping websites, how they handled these issues, and their suggestions for improvements. In the second part, participants were asked to \textit{``screen share and think aloud''} while browsing a new shopping website, Mercari. We selected Mercari because it is a legitimate, English-language shopping platform that is not widely known among our participant group. This helped ensure that participants were unfamiliar with the site, allowing us to better observe their natural exploration behaviors and minimize prior experience bias in evaluating accessibility. We instructed them to start by browsing the homepage and exploring the website, then presented them with a scenario of searching for and selecting a blender on the website. Participants were asked to think aloud about the keys they used and their thoughts on the browsing experience with screen readers while performing these tasks. %Data analysis details are in section~\ref{sec:data_ana}.

\subsection{Data Analysis}
All interview recordings from our formative study were transcribed using Zoom. We employed a deductive thematic analysis approach~\cite{braun2006using}, guided by our initial research questions such as identifying accessibility challenges on unfamiliar shopping websites, but we also remained open to new, inductively identified themes. Three authors independently reviewed two randomly selected transcripts, writing down preliminary codes that captured both explicit statements (for example, ``lack of comprehensive alt text'') and implicit ideas (for example, ``user frustration with new page layouts''). After this initial pass, the three coders met to reconcile differences, merge overlapping ideas, and refine code definitions into a single preliminary codebook. Each coder then applied this refined codebook to the remaining transcripts, continuing to add or revise codes when new concepts emerged. Throughout the coding process, all coders meet regularly to resolve any discrepancies by discussing how best to interpret particular quotes in light of the existing codebook. This process yielded around 30 themes, such as ``Accessibility Issues from Improper HTML Tags'' and ``Preference for Independent Browsing,'' which were grouped under broad categories like ``Navigation Barriers,'' ``Workarounds,'' and ``Desired Enhancements.''

\subsection{Interview Findings}
% \subsubsection{Current Challenges and Practices}
All participants encountered accessibility barriers when using online shopping websites. These barriers included inappropriate use of HTML tags and labels, a lack of comprehensive alt text for images, and an offline return process. In the following sections, we will discuss each of these challenges in detail and report the strategies participants used to counter them.

\subsubsection{Current Accessibility Challenges on Shopping Websites}

\paragraph{\textbf{\textit{Inappropriate Use of HTML Tags and Labels.}}}
All participants highlighted  the inappropriate use of HTML tags and labels as the most common accessibility issue on shopping websites. Screen reader users reported relying on shortcut keys to navigate HTML components efficiently. P1 explained, \textit{``Most of the time, if I enter a website, I press H to check if they have headings or not, because this is the fastest and easiest way for us to browse.''} Common shortcuts include pressing "H" to iterate through all headings or numbers 1 to 6 to navigate specific heading levels ~\cite{yu2023design, kaushik2023guardlens, webaim_screen_reader_survey_10}. Misuse of HTML tags like headings frustrates users with excessive, irrelevant information, making it hard to locate key details or understand crucial information needed for purchase decisions.

% All participants highlighted that the most prevalent accessibility issue on shopping websites is the inappropriate use of HTML tags and labels. Screen reader users reported that they typically use shortcut keys to quickly navigate between specific HTML components based on their tags. P1 explained in the interview, \textit{``Most of the times if I enter website, I press H to check if they have headings or not, because this is the fastest and the easiest way for us to browse.''} The most commonly used shortcuts involve jumping between headings ~\cite{yu2023design, kaushik2023guardlens, webaim_screen_reader_survey_10}: pressing ``H'' allows users to sequentially iterate over all headings regardless of their level, while pressing numbers 1 to 6 enables them to browse through headings of a specific level one at a time. Inappropriate use of HTML tags such as heading leads to screen reader users' frustration from hearing excessive unnecessary information, difficulty in locating key details, and challenges in understanding crucial information needed for making purchase decisions.

\paragraph{\textbf{Too many headings.}}
%P5 "And then I search, and then I press heading. I'm assuming the search results are heading by headings. Yes, this is the heading filter, by oh, 1,000 results or so."
%P2: "00:56:48.000 There's just like so many headings. Felt like.Unnecessary like. I would have liked.

Five out of six participants reported that shopping websites often feel overcrowded due to excessive emphasis on non-essential information through headings, causing confusion about the website's structure and flow, particularly during initial browsing. Overuse of headings makes it hard for screen reader users to navigate quickly, as they must listen to irrelevant content. For instance, some websites place long lists of category headings, like electronics or home, before the main content, forcing users to navigate through numerous headings—often over ten—before reaching key information. P2 noted, \textit{``Categories don’t have to be headings; there could just be a heading titled `Categories,' with items listed as links underneath.''} Similarly, P1 mentioned, \textit{``Sometimes all the filters are marked as headings, making it take a long time to reach the product search results.''}
% Many participants (5 out of 6) reported that shopping websites often feel overcrowded due to excessive emphasis on non-essential information through headings, which creates confusion for screen reader users regarding the website's structure and flow. This issue is particularly problematic during their initial browsing experience. The overuse of headings makes it difficult for screen reader users to quickly navigate to key information, as they are forced to listen to irrelevant content. For example, some websites place long lists of category headings, such as electronics or home, before the main content. As a result, screen reader users must navigate through numerous category headings—often more than ten—before reaching the essential information they are interested in, significantly increasing navigation time and effort. P2 specifically pointed out that some websites unnecessarily use headings for every category, \textit{``Categories do not have to be headings; there could just be a heading titled ``Categories'', with each item listed as a link underneath.''}  P1 also mentioned that \textit{``Sometimes all the filters are marked as headings, which makes it take a long time to reach the product search results.''}

\paragraph{\textbf{Lack of headings.}}
%P4: So, right? So on mobile, it's always been the same, like, you know, like, for example, like, swipe right, it tells me the button or heading or whatever. Like, headings are usually more for there's not, you know, something here to enter. So there's not as many headings on mobile, this might be buttons or links.
On the contrary, some informal yet essential information is not placed in headings or appropriate HTML tags, misaligning with screen reader users' navigation preferences. Screen reader users typically jump between headings, links, and buttons to quickly access key information ~\cite{yu2023design,webaim_screen_reader_survey_10}. P3 explained, \textit{``When I am searching for something, I really expect it to be in a heading.''} For example, some shopping websites fail to include product titles in headings. Participants often attempted to navigate by headings to locate products but, failing to do so, resorted to using the down arrow key to iterate through elements on the search results page. P4 noted, \textit{``Some smaller store websites are not accessible at all; they are too graphical with not enough headings and links to navigate.''}

% other informal and essential information sometimes is not in headings or other appropriate HTML tags that align with screen reader users' browsing preference. This mismatch does not align with the way screen reader users typically navigate by jumping between headings, links, and buttons to quickly access the most important information ~\cite{yu2023design,webaim_screen_reader_survey_10}. P3 explained in the interview, \textit{``when I am searching for something, I am really expecting that it will be in heading.''}  For example, some shopping websites do not include the product title in the headings. Participants found first navigated by headings to find the products and failed, so they need to use down arrow key to iterate each element to locate where the products start on the search result page. P4 explicitly mentioned \textit{``Some of the smaller store websites are not accessible at all; they are too graphical with not enough headings and links to navigate.''} 

\paragraph{\textbf{Disorganized Heading Hierarchy.}}
% P4: So escape. You know, it takes it back to the top of the page every time. So once I get in, you know, when I pick my selection, then it goes take me back to the top, and then I have to do h again, popular searches, heading level six, TV, remote. And again, I'm arrowing down. So I did H couple of times. Now I'm arrowing down once, heading up any for so, so at all, like one, TV, one, tell one, dad, five, cluster three. DVD, one.
Even when appropriate headings are used, inconsistent heading levels and poorly structured layouts can still pose challenges for screen reader users. For instance, a website might assign a level 1 heading to reviews and a level 6 heading to product descriptions, confusing users who rely on specific heading levels for navigation.

Many participants navigate headings using the ``H'' key, following their sequence in the HTML code rather than their visual layout. This often leads to a disjointed experience, as visually oriented layouts may prioritize sighted users. For example, if the layout places images and reviews on the left and product names and details on the right, screen reader users will hear about reviews and vague image descriptions first, instead of the crucial product names and details. Screen reader users may encounter irrelevant information, like reviews or vague image descriptions, before crucial product details. P5 noted, \textit{``I just don’t understand why unnecessary information like likes, comments, and share buttons is positioned before the product information.''}

% Even when shopping websites use appropriate headings for content, the assignment of heading levels and the layout of these elements can still be challenging for screen reader users. For example, a website might use headings for seller information, reviews, and product descriptions but inconsistently assign heading levels, such as using a level 1 heading for reviews and a level 6 heading for the product description. This inconsistency can confuse users who rely on navigating by specific heading levels. 

% Many participants use the ``H'' key to navigate through headings in the order in which they appear in the HTML code, without considering the heading levels. This can lead to a disjointed browsing experience if the visual layout of the page is designed primarily for sighted users. For example, if the layout places images and reviews on the left and product names and details on the right, screen reader users will hear about reviews and vague image descriptions first, instead of the crucial product names and details. This mismatch in the layout and navigation order creates confusion and makes it harder for screen reader users to find key information efficiently. P5 explicitly mentioned the example and noted \textit{``I just do not understand why unnecessary information like likes, comments, and share buttons is positioned before the product information.''}

\paragraph{\textbf{Unclear Labels on Images and Buttons.}}
Participants reported frustration with the lack of clear, descriptive labels for buttons and images, a common issue on less mainstream shopping websites. Unclear labels confuse screen reader users, who rely on meaningful descriptions to navigate interactive elements. For instance, a button might be labeled generically as ``button'', providing no indication of its function, such as ``add to cart'' or ``view details.'' Similarly, images may lack alt text, leaving their content or relevance unclear. This forces users to guess button functions or spend unnecessary time exploring, reducing usability and accessibility issues. P2 noted, \textit{``Amazon has a few unlabeled buttons that can be a bit confusing sometimes. You can usually figure them out from the context, but they are not labeled correctly.''}

% Participants reported frustration with the lack of clear and descriptive labels for buttons and images, a common issue on less mainstream shopping websites. Unclear labels create confusion and hinder navigation for screen reader users who rely on meaningful text descriptions to understand the function of interactive elements. For example, a button might be generically labeled as the ``button'' without any additional context, making it impossible for users to know what action it performs, such as ``add to cart'' or ``view details.'' Similarly, images may lack descriptive alt text, preventing users from understanding their content or relevance. This lack of clarity forces screen reader users to guess the purpose of buttons and images or to spend time exploring them unnecessarily, significantly diminishing the overall usability and accessibility of the website. Without meaningful labels, these users cannot efficiently navigate the site or complete tasks, which can lead to frustration and abandonment of the site altogether. For example, P2 mentioned that \textit{``Amazon has a few unlabeled buttons, that can be a bit confusing sometimes. You can usually figure them out from the context, but they are not labeled correctly.''}

\paragraph{\textbf{\textit{Lack of comprehensive and accurate text and image description for products.}}}
Participants noted difficulties understanding product details on shopping websites due to insufficient text and image descriptions. Many sites rely heavily on images to convey key information, such as color, style, condition, and features, without providing adequate text. This creates a significant barrier for screen reader users, who depend on alt text to interpret images. P2 shared, \textit{``The condition of items on eBay is hard to identify because it's all about images.''} Participants also mentioned inconsistencies between product descriptions and reviews, adding to the confusion. P3 explained, \textit{``When it comes to clothing, I want more detail than just `blue shirt.' I need to know about the fit and style, which is often missing in online descriptions.''}

% Participants highlighted a challenge in trying to understand product details on shopping websites due to a lack of comprehensive text and image descriptions. Many websites often rely heavily on images to convey important information about products, such as color, style, condition and specific features, without providing adequate accompanying text descriptions. This practice poses a major accessibility barrier for screen reader users who cannot see these images and instead rely on alt text to understand the content. For example, P2 shared that \textit{``The condition of items on eBay is hard to identify because it's all about images.''} The participants also shared that product descriptions sometimes contradict product reviews, creating confusion. They often need to verify details shown in images, but the lack of clear and comprehensive image descriptions makes this verification difficult. P3 shared that \textit{``When it comes to clothing, I want more detail than just ``blue shirt.'' I need to know about the fit and style, which is often missing in online descriptions.''}

\paragraph{\textbf{\textit{Hard to compare different products.}}}
Three out of six participants reported significant challenges comparing similar products on shopping websites. While sighted users can quickly assess visual and descriptive information, screen reader users must repeatedly switch between tabs to find key details like prices, descriptions, and specifications. This process is time-consuming and mentally exhausting. P3 explained, \textit{``Comparing products, especially similar ones, is just exhausting. I have to remember all the details in my head and switch back and forth between tabs to check things like the price or specific features.''} Improper use of HTML tags and labels further complicates navigation, making the task even more inefficient. P2 similarly noted, \textit{``It’s really hard to compare products that are very similar. I have to keep switching between tabs to find the price or some specific details. It’s not efficient at all.''}
% Some participants (3 out of 6) reported significant challenges when comparing different products, particularly similar models, on shopping websites. While sighted users can quickly compare visual information and descriptive text, screen reader users must repeatedly switch between tabs to locate key details, such as prices, descriptions, and product specifications. This appears to be a unique challenge for screen reader users that was relatively unexplored in previous literature. Users report that the process is not only time-consuming, but also mentally exhausting. For example, P3 explained, \textit{``Comparing products, especially similar ones, is just exhausting. I have to remember all the details in my head and switch back and forth between tabs to check things like the price or specific features.''} The issue is further compounded by the improper use of HTML tags and labels on these websites, which makes navigation even more cumbersome. P2 shared a similar sentiment, stating, \textit{``It’s really hard to compare products that are very similar. I have to keep switching between tabs to find the price or some specific details. It’s not efficient at all.''}

\subsubsection{Strategies for Overcoming Accessibility Challenges}

\paragraph{\textbf{\textit{Relying on External Video Reviews.}}
\normalfont{Participants reported relying on external resources, such as YouTube reviews, to counter accessibility challenges and gain a clearer understanding of products. These video reviews provide in-depth demonstrations and visual details that are often lacking in the brief text descriptions found on shopping sites. For example, P4 mentioned, }\textit{``When I’m not sure about how a product looks or functions, I usually check YouTube. The videos give me a much better idea than the short descriptions on the website.''} \normalfont{This approach helps participants make more informed decisions, especially when considering the purchase of products where visual details are important, such as electronics or clothing.}}

\paragraph{\textbf{\textit{Utilizing User Ratings and Reviews}}
\normalfont{Additionally, participants closely follow user ratings and reviews on shopping platforms to gauge the condition and quality of products. Since product images or brief descriptions can be misleading or incomplete, user reviews become a crucial source of information. These reviews often provide first-hand experiences and specific details that are not immediately visible or described in the official product listings. As P2 explained, }\textit{``The product pictures don’t always tell the full story. I go through the reviews to see what people are actually saying about the product condition and quality.''}}

\paragraph{\textbf{\textit{Seeking Assistance from Sighted Individuals}}
\normalfont{Participants might ask sighted friends or family members for assistance in describing visual details that are important for making purchase decisions. This practice is necessary when the product’s visual information is critical but inaccessible. P5 shared their experience, stating,} \textit{``Sometimes I have to ask someone to look at it to make sure I am seeing the right thing, especially when it is not clear on the website.''} \normalfont{This indicates a reliance on others when the digital content is insufficient, which adds another layer of dependency and effort for screen reader users.}}

\paragraph{\textbf{\textit{Leveraging GenAI for Accessibility Enhancements}}
\normalfont Interestingly, most of the participants (4 out of 6) have used GenAI to help mitigate accessibility challenges on shopping websites. They reported using GenAI tools such as ChatGPT ~\cite{chatgpt} and Be My Eyes ~\cite{bemyeyes} to get detailed descriptions of product images and clarify ambiguous information that was not mentioned in the text description or even in the image alt text. P5 shared that \textit{``I used ChatGPT to describe pictures of the product and got a better idea of what the product consists of. It would be great to have an AI that can answer questions about how a product looks.''} \normalfont{Participants also used GenAI to compare different models of a product by generating a navigable table that highlights key differences, making it easier to understand these differences and make informed purchase decisions}. P6 also shared, \textit{``I usually make the AI compare two products by giving it links or screenshots of the pictures to find the differences that are not mentioned in the descriptions.''} \normalfont{However, they also reported limited trust in AI to conduct product searches because it often does not bring up results from preferred websites such as Amazon. P6 explained,} \textit{``I do not trust AI to search for products on Amazon. I usually search for the product myself and then use AI to make sure it has the features I am looking for.''}}

\paragraph{\textbf{\textit{Preferred Improvements for Screen Reader Users}}
\normalfont{Participants preferred browsing websites themselves over using virtual assistants, which have been proposed as potential solutions~\cite{virtual_assistant_vtyurina2019verse}. They cited concerns about reliability, language compatibility, and additional accessibility barriers. P1 explained, \textit{``I prefer browsing myself because I can control what I do with my screen reader.''} P2 noted issues with virtual assistants, stating, \textit{``If you are in the same window with a virtual assistant, you are not aware of new messages. That is not friendly for screen reader users at all.''} Instead, participants suggested improvements like a screen reader mode'' that simplifies web pages by showing only essential elements, such as product descriptions. P3 shared, \textit{``It would be helpful if shopping websites had a screen reader mode that strips away images and just shows the product descriptions.''} This highlights their preference for streamlined, user-controlled experiences that focus on relevant information.}}

\subsection{Design Implications}
Our findings strongly support our design intuition that leveraging GenAI to improve HTML tags and labels can significantly enhance accessibility for screen reader users. Participants already use GenAI to tackle certain accessibility challenges, such as extracting detailed descriptions from images or comparing similar products. However, directing GenAI to effectively revise the layout and information hierarchy through HTML tags is challenging and often unfeasible for them. This difficulty highlights the need for more intuitive, user-friendly solutions that seamlessly integrate with existing browsing practices, reducing the burden on users to manually adjust or work around accessibility issues.

Participants expressed a preference for accessibility solutions that operate within their familiar browsing environments rather than relying on external tools that add layers of complexity, such as AI assistants designed to help locate information. This suggests a need for back-end solutions where developers revise a website’s structure to better align with how screen reader users navigate, creating a more seamless and integrated user experience. However, this is a challenging task for developers, as it requires specialized knowledge of both accessibility standards and screen reader behavior, along with substantial time and effort to implement these changes manually.

We developed a GenAI-driven tool to enhance online shopping website accessibility by optimizing HTML tags, labels, and information hierarchy. The system addresses common barriers for screen reader users, such as inconsistent headings, missing alt text, and unclear button labels. By automating these adjustments, the tool reduces developer effort and bridges knowledge gaps, offering a more efficient way to create accessible web experiences.

\section{System Implementation}
\label{sec:phase2}

\subsection{System Overview and Purpose}
To address the accessibility challenges identified in our formative study, we proposed the concept of using Generative AI (GenAI) to dynamically restructure webpage content for screen reader users. While participants identified a lack of descriptive image labels as an issue, we excluded image description generation, as participants and previous research already used GenAI to address this independently \cite{10.1145/3234695.3236337, huh2023genassist}. Our approach aims to improve navigation efficiency, logical content hierarchy, and labeling of interactive elements. To test this concept, we developed a browser extension as a proof-of-concept system to evaluate its feasibility and gather user feedback on preferences and experiences.

The system is designed to operate in two modes: %\textbf{(1)} rewriting the entire website, or \textbf{ (2)} only modifying the tag structure but leaving the website otherwise unmodified.

\begin{enumerate}
    \item \textbf{Rewrite the Document: Regenerated HTML Version (Option 1). } In this option, we produce an HTML document which is re-created from scratch to provide an accessible screen reader experience. The content is preserved, but all other aspects of the site can be modified. The elements can be re-ordered, and visual design aspects may be removed, since the usability of the screen reader is the only criteria under consideration.
    \item \textbf{Replace the Tags: Reorganized HTML Tags Version (Option 2).} The second option adjusts existing tags without altering the visual layout, focusing on minimal changes to preserve content integrity while improving navigability. This goal is achieved by re-writing part of the website's Document Object Model (DOM) to reorganize HTML tags, fix misused elements, and provide better labels.
\end{enumerate}

\subsection{Technical Architecture} 
\begin{figure}[h]
    \centering
    \includegraphics[width=1\linewidth]{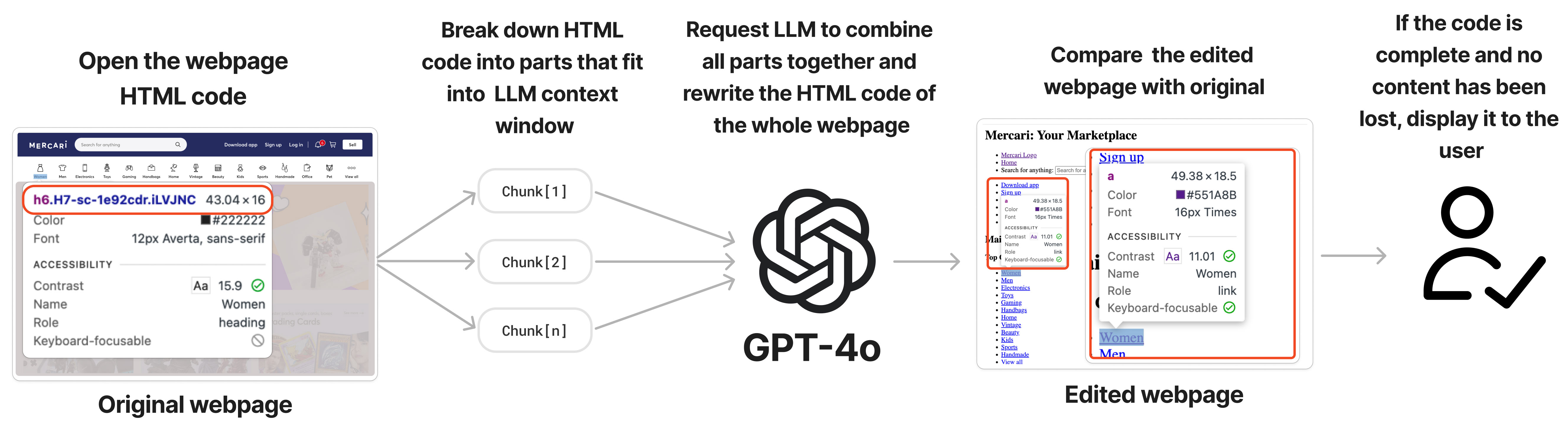}
    \caption[This diagram illustrates the process of using GPT-4o to rewrite HTML code for improved accessibility. The process flows from left to right, starting with opening the webpage HTML code, represented by a product page showing items like cameras, headphones, and smartwatches with prices. The HTML is then broken down into multiple chunks that fit within the LLM's context window, labeled as ``Chunk\[1\]'' to ``Chunk\[n\]''. These chunks are fed into GPT-4o, depicted by a stylized brain icon, which combines all parts and rewrites the HTML code for the entire webpage. The final step shows the output webpage, visually similar to the original, but potentially restructured for better accessibility. Arrows connect each stage, illustrating the progression from the initial webpage through the GPT-4o processing to the final, accessibility-enhanced output. The diagram emphasizes the seamless transformation of web content for improved screen reader compatibility while maintaining visual consistency.]{\fixme{Option 1 pipeline overview, prompting GPT-4o model to completely rewrite HTML code}}
    \label{fig:1}
\end{figure}

\begin{figure}[h]
    \centering
    \includegraphics[width=1\linewidth]{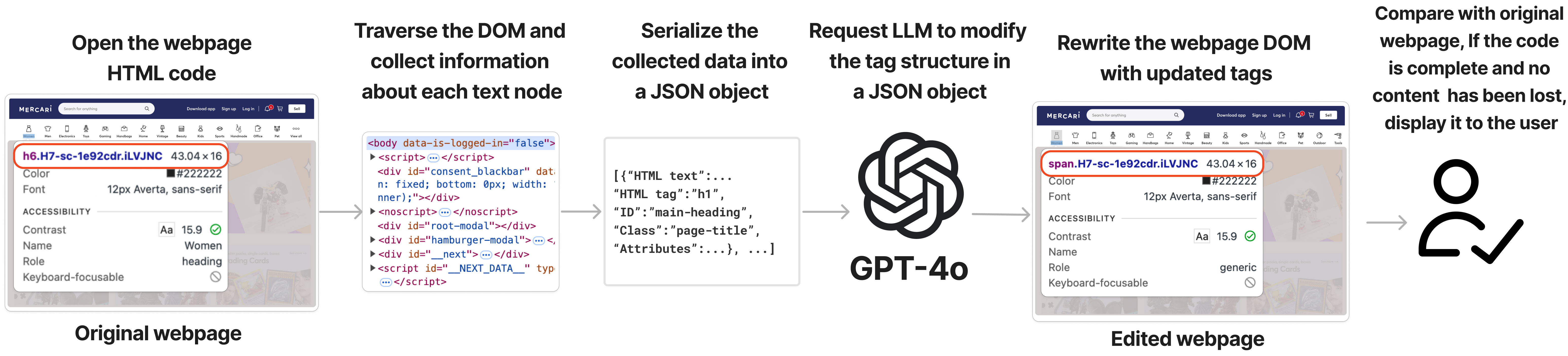}
    \caption[This diagram illustrates a process for improving web accessibility using GPT-4o, progressing from left to right through six stages. It begins with opening the webpage HTML code, shown as a product page displaying items like cameras and headphones. The next step involves traversing the Document Object Model (DOM) and collecting information about each text node, represented by a snippet of HTML code. This collected data is then serialized into a JSON object, containing details such as HTML text, tags, IDs, classes, and attributes. The JSON object is fed into GPT-4o, symbolized by a stylized brain icon, which modifies the tag structure. Following this, the webpage's DOM is rewritten with the updated tags. The final step compares the modified version with the original webpage, displaying the result to the user if the code is complete. The output appears visually similar to the input, suggesting that the accessibility improvements maintain the page's visual integrity while enhancing its structure for screen readers. Arrows connect each stage, depicting the flow from the initial webpage through the GPT-4o processing to the final, accessibility-enhanced output.]{\fixme{Option 2 pipeline overview, prompting GPT-4o to rewrite only some parts of the HTML code}}
    \label{fig:2}
\end{figure}

We developed a Google Chrome Extension and tested it in Google Chrome Official Build 127.0.6533.122 (arm64). Google Chrome was the most widely-used browser and had largest number of screen readers deployed in a desktop context \cite{webaim_screen_reader_survey_10}. The pipeline of our system is applicable to other browsers and computing environments, including mobile web browsers that support plugins. We implemented the prototype leveraging Chrome’s Web APIs for real-time DOM access and manipulation. The extension extracts webpage content, processes it with GPT-4o via the OpenAI API, and reinserts the modified HTML or tag updates back into the live webpage. We selected GPT-4o for its support of large input/output sizes (128K/16K tokens), which is critical for processing full HTML pages. The system processes content in segments, preserving coherence through sequential prompting and chunk-level context injection.
\begin{itemize}
    \item \textbf{Option 1 (see Figure~\ref{fig:1})} shows how the system decomposes a webpage into chunks that fit the model’s context window, sends these chunks to GPT-4o along with prior context, and reconstructs the full page by stitching outputs together. 
    \item \textbf{Option 2 (see Figure~\ref{fig:2})} follows a similar pipeline but focuses on tag-level editing instead of complete rewriting. The model receives structured segments and returns modified HTML tags, which are directly updated in place. This method preserves the original visual layout and reduces the risk of content discrepancies.
\end{itemize}
We configured the LLM with deterministic parameters (\textit{temperature = 0.2}, \textit{max\_tokens = 16,384}, \textit{top\_p = 1}, \textit{frequency\_penalty = 0}, and \textit{presence\_penalty = 0}) to ensure reproducibility. Furthermore, we also tried our prototype with similar models such as Google Gemini 1.5 Pro \cite{gemini} and Anthropic Claude 3.5 Sonnet \cite{claude}, and found that in all cases the models could generate valid HTML code consistent with the intended prompt. Therefore, we believe that our approach can be sufficiently robust in multiple advanced LLMs. We ultimately chose GPT-4o for our implementation due to its strong performance, wide availability, and efficient integration with our existing development environment. Additional implementation details and prompt templates are available in Appendix~\ref{appendix}.

To illustrate the types of enhancements made under both versions, Figure~\ref{fig:teaser} presents a before-and-after example showing how HTML is transformed for improved accessibility for both versions. While Option 1 applies broader restructuring to the entire page, Option 2 preserves layout while correcting semantic structure. A summary of representative changes for each version is provided in Appendix~\ref{appendix:examples}.

\subsubsection{Mitigating Content Discrepancies in LLM-Generated Accessibility Enhancements}
Large language models (LLMs) can effectively revise and improve accessibility design; however, they are prone to content discrepancies~\cite{8496795}, where certain details may be unintentionally omitted or altered when regenerating a webpage. In our context, this can manifest as missing content (e.g., product links) or unintended text in the HTML output. To address these challenges, we adopted two key strategies. First, we refined prompts and implemented a similarity check function to identify discrepancies between the regenerated page and the original. When discrepancies were detected, such as missing product links on extensive pages like Amazon’s search results, we programmatically reinserted these details to ensure consistency with the source. The LLM-generated result was only rendered on the front end if the similarity measure exceeded a threshold of 90\%. This threshold was chosen based on preliminary tests, which showed that values below 90\% often reflected differences due to reformatting rather than critical content changes. Second, we limited some revisions to HTML tags only in our Option 2 approach, which minimized the risk of losing key information but also reduced our ability to reorder visual layouts for better accessibility (see Section~\ref{sec:phase3}).

\vspace{-0.14in}

\section{System Evaluation}
\subsection{Automatic Accessibility Evaluation}
\label{auto_eval}
To assess whether our GenAI-enhanced webpages compliy with WCAG guidlines, we conducted automatic audit using three widely adopted accessibility testing tool, Google Lighthouse~\footnote{https://developer.chrome.com/docs/lighthouse/overview}, SortSite~\footnote{https://www.powermapper.com/products/sortsite/}, and AChecker~\footnote{https://websiteaccessibilitychecker.com/checker/index.php}. We ran each tool on the original webpages and their LLM-regenerated version and LLM-reorganized version. For the evaluation, we chose three e-commerce websites—Amazon, Nordstrom, and Mercari—based and tested three common pages: Home page, Search Results Page, and Product Page. This selection was informed by user interviews in the formative study, which provided insights into platform familiarity and usage patterns. Specifically, we included: a widely used platform (e.g., Amazon), a platform that users recognized but rarely used (e.g., Nordstrom), and a platform that was largely unfamiliar (e.g., Mercari). This range allowed us to capture a variety of accessibility challenges experienced by screen reader users across websites with differing levels of user familiarity and interface complexity. As shown in Table~\ref{tab:accessibility_problems_double}, both the LLM-regenerated and LLM-reorganized versions consistently reduced the number of accessibility violations across all three tools. For example, on the Amazon Product Page, Google Lighthouse reported a drop from 16 violations in the original to one in the regenerated version and three in the reorganized version. Similarly, SortSite reported five violations on the Nordstrom Home Page in the original, which dropped to one in the regenerated version and one in the reorganized version. Across all platforms and page types, the regenerated version showed the most substantial improvements, particularly in reducing issues related to heading hierarchy, ARIA roles, and unlabeled interactive elements.

\begin{table}[h]
\centering
\renewcommand{\arraystretch}{1.2} % Adjust row height for better readability
\resizebox{\textwidth}{!}{%
\begin{tabular}{llccc|llccc}
\hline
\textbf{Page} & \textbf{Category} & 
\textbf{Lighthouse} & 
\textbf{SortSite} & 
\textbf{AChecker} & 
\textbf{Page} & \textbf{Category} & 
\textbf{Lighthouse} & 
\textbf{SortSite} & 
\textbf{AChecker} \\ \hline

\textbf{Mercari Home Pages} & Original & 6 & 4 & 6 & 
\textbf{Nordstrom Home Pages} & Original & 3 & 5 & 3 \\
                              & Option 1 & 1 & 0 & 0 & 
                              & Option 1 & 1 & 1 & 0 \\
                              & Option 2 & 2 & 2 & 3 & 
                              & Option 2 & 1 & 1 & 1 \\ \hline

\textbf{Mercari Search Result Pages} & Original & 5 & 5 & 6 & 
\textbf{Nordstrom Search Result Pages} & Original & 3 & 3 & 3 \\
                                       & Option 1 & 1 & 0 & 1 & 
                                       & Option 1 & 1 & 1 & 0 \\
                                       & Option 2 & 1 & 2 & 2 & 
                                       & Option 2 & 2 & 1 & 1 \\ \hline

\textbf{Mercari Product Pages} & Original & 6 & 4 & 5 & 
\textbf{Nordstrom Product Pages} & Original & 2 & 6 & 3 \\
                                 & Option 1 & 1 & 1 & 1 & 
                                 & Option 1 & 1 & 1 & 0 \\
                                 & Option 2 & 2 & 2 & 3 & 
                                 & Option 2 & 1 & 3 & 1 \\ \hline

\textbf{Amazon Home Pages} & Original & 6 & 6 & 7 & 
\textbf{Amazon Search Result Pages} & Original & 12 & 17 & 13 \\
                                   & Option 1 & 1 & 1 & 2 & 
                                   & Option 1 & 1 & 0 & 1 \\
                                   & Option 2 & 2 & 2 & 3 & 
                                   & Option 2 & 3 & 4 & 3 \\ \hline

\textbf{Amazon Product Pages} & Original & 16 & 18 & 15 &\\ 
                               & Option 1 & 1 & 1 & 1 \\
                               & Option 2 & 3 & 4 & 4 \\ \hline

\end{tabular}%
}
\caption{\fixme{Comparison of Level A accessibility issues detected on various webpages using three automated evaluation tools: Lighthouse, SortSite, and AChecker (lower is better). A lower number indicates improved accessibility. The table shows the number of accessibility problems identified for the original version of each page, Regenerated HTML Version (Option1), and Reorganized HTML Tags Version (Option2). While Option 1 achieved near-perfect results in most cases, minor issues like one/two broken links persisted. Option 2 addressed structural problems but retained challenges such as inaccessible heading sequences and visual design conflicts due to its inability to alter the original layout or content order. 
% \   {maybe briefly summarize the kinds of accessibility issues that the current version of option 1/2 still have. those need to elaborated in the discussion and suggest for future work}
}}
\label{tab:accessibility_problems_double}
\end{table}

%\vspace{-0.35in}
\subsection{\fixme{Content Integrity Evaluation}}
\label{sec:evaluation}
While improving accessibility, it is equally important that the regenerated webpages preserve the original content. To evaluate this, we measured Aggregated Semantic Similarity between the original and LLM-generated versions. We selected this metric because it aligns with the real-world experience of screen reader users, and is recommended in Microsoft’s AI evaluation guidelines~\cite{microsoft2023evaluation} and supported by prior research~\cite{chen2021evaluating}. This metric captures whether the meaning and intent of text within HTML elements remain consistent, even when phrasing or structure changes.
% \   {is this a common metrics used in web accessibility research? If so, I’d suggest adding a citation. }

To ensure meaningful alignment, we extracted all screen-reader-accessible HTML elements, such as visible text, ARIA labels, and alternative text, from both the original and regenerated versions, and compared their aggregated semantic embeddings. Unlike string-level comparisons, this metric accounts for paraphrasing, content reordering, and summarization—which are common outcomes of LLM-based regeneration. For instance, moving ``Product Details'' above ``Customer Reviews'' or collapsing redundant filter headings may improve usability while maintaining semantic equivalence.

We applied this evaluation to the same three websites and page types used in the accessibility audit. Since Option 2 only modifies HTML tags without changing content, we focused the analysis on Option 1 (Regenerated HTML). As shown in Table~\ref{tab:match_quality}, the regenerated pages achieved high semantic similarity scores, frequently exceeding 95\%, indicating strong content preservation. In a few cases where the scores were slightly lower, such as the Nordstrom homepage, we observed that the LLM had condensed or reorganized content—for example, grouping scattered promotional items or combining fragmented text elements into single, coherent sections. These changes slightly reduced the similarity score, but reflected practical adjustments that enhance clarity and navigation for screen reader users.

\begin{table}[h]
\centering
\resizebox{0.7\textwidth}{!}{%
\begin{tabular}{lcc}
\hline
\textbf{Platform} & \textbf{Page Type} & \textbf{Aggregated Semantic Similarity (\%)} \\ \hline
Mercari           & Home Page          & 99.36 \\
Mercari           & Search Results Page & 97.09 \\
Mercari           & Product Page       & 98.35 \\
Nordstrom         & Home Page          & 91.60 \\
Nordstrom         & Search Results Page & 93.19 \\
Nordstrom         & Product Page       & 92.06 \\
Amazon            & Home Page          & 97.47 \\
Amazon            & Search Results Page & 98.56 \\
Amazon            & Product Page       & 98.89 \\ \hline
\end{tabular}%
}
\caption{Aggregated Semantic Similarity scores between the original and LLM-regenerated webpages across three e-commerce platforms (Amazon, Nordstrom, and Mercari) and three page types (Home, Search Results, and Product). These results demonstrate strong preservation of content meaning, even when structure and phrasing were altered to improve accessibility. Lower scores, such as on the Nordstrom homepage, reflect cases where the LLM reorganized or merged content for better screen reader usability.}
\label{tab:match_quality}
\end{table}

% \vspace{-0.14in}

\vspace{-0.35in}
\section{User Evaluation}
\label{sec:phase3}
While automated audits provide useful indicators of accessibility, prior research cautions against relying solely on these tools due to inconsistencies in their coverage and reliability \cite{benchmarking}. To complement our technical evaluation and better understand the real-world impact of our system, we conducted a user evaluation with screen reader users. We focused the user evaluation on Mercari.com because it is a legitimate but unfamiliar English-language shopping site, which helps minimize prior experience bias. In addition, since each session lasted up to 90 minutes, we limited testing to one website to avoid overburdening participants and to allow for in-depth comparison of the three versions.

\subsection{Participants Demographics}
\fixme{15} screen reader users participated in the study, consisting of \fixme{nine} females and \fixme{six} males. The participants' ages ranged from 18 to 44 years. All demographic data were self-reported by the participants (see Table \ref{tab:phase2demo}). \fixme{Among the participants, 11 were completely blind, one had limited vision, and three could perceive only light.} Participant recruitment details are in section \ref{sec:recruit}.

% recover
\begin{table}[h]
\resizebox{1\textwidth}{!}{
\begin{tabular}{llllll}
\hline
\textbf{ID} &
  \textbf{Age} &
  \textbf{Gender} &
  \textbf{Visual Ability} &
  \textbf{\begin{tabular}[c]{@{}l@{}}Online shopping\\ Frequency\end{tabular}} &
  \textbf{\begin{tabular}[c]{@{}l@{}}Used Screen\\ Reader\end{tabular}} \\ \hline
\textbf{P1} & 25 - 34 & Female & I am totally blind                & 4-6 times a week & JAWS, NVDA, VoiceOver       \\
\textbf{P2} & 25 - 34 & Female & Completely blind                  & 4-6 times a week & JAWS                        \\
\textbf{P3} & 25 - 34 & Female & Blind with light perception & Once a week      & JAWS, NVDA, Google Talkback \\
\textbf{P4} & 18 - 24 & Female & No vision                         & Once a week      & JAWS, NVDA, VoiceOver       \\
\textbf{P5} & 35 - 44 & Female & Totally blind                     & Once a week      & JAWS, NVDA                  \\
\textbf{P6} & 25 - 34 & Male   & Blind, light perception only      & 2-3 times a week & JAWS, NVDA                  \\
\textbf{P7} & 25 - 34 & Male   & I have limited vision             & 2-3 times a week & JAWS                        \\ 
\fixme{\textbf{P8}} & \fixme{25 - 34} & \fixme{Male}   & \fixme{Fully blind}  & \fixme{4-6 times a week} & \fixme{JAWS, NVDA, Google Talkback, VoiceOver}  \\ 
\fixme{\textbf{P9}} & \fixme{35 - 44} & \fixme{Female}   & \fixme{Legally blind}  & \fixme{4-6 times a week} & \fixme{JAWS, Google Talkback, VoiceOver}  \\ 
\fixme{\textbf{P10}} & \fixme{35 - 44} & \fixme{Male}   & \fixme{Blind}  & \fixme{Once a week} & \fixme{JAWS, VoiceOver}  \\ 
\fixme{\textbf{P11}} & \fixme{25 - 34} & \fixme{Female}   & \fixme{Completely blind}  & \fixme{2-3 times a week} & \fixme{JAWS, NVDA, Google Talkback, VoiceOver}  \\ 
\fixme{\textbf{P12}} & \fixme{25 - 34} & \fixme{Female}   & \fixme{Zero vision}  & \fixme{Once a week} & \fixme{JAWS, NVDA, Google Talkback, VoiceOver}  \\ 
\fixme{\textbf{P13}} & \fixme{25 - 34} & \fixme{Female}   & \fixme{Blind but can perceive light}  & \fixme{Once a week} & \fixme{JAWS, NVDA, Google Talkback, VoiceOver}  \\ 
\fixme{\textbf{P14}} & \fixme{35 - 44} & \fixme{Male}   & \fixme{Legally blind}  & \fixme{2-3 times a week} & \fixme{JAWS, VoiceOver}  \\ 
\fixme{\textbf{P15}} & \fixme{25 - 34} & \fixme{Male}   & \fixme{Blind}  & \fixme{Once a week} & \fixme{JAWS, Google Talkback}  \\ 
\hline
\end{tabular}
}
\caption[Table 2. User Evaluation Participant Demographic Table, which details the characteristics of seven participants (P1 through P7) in a user evaluation study. The table consists of six columns: ID, Age, Gender, Visual Ability, Online shopping Frequency, and Used Screen Reader. The participants are predominantly female (5 out of 7) with ages ranging from 18-44, mostly in the 25-34 range. All participants have some form of visual impairment, from total blindness to limited vision. Online shopping frequency varies from once a week to 4-6 times a week. Most participants use multiple screen readers, with JAWS being the most common, followed by NVDA and VoiceOver. This table provides a comprehensive overview of the diverse group of visually impaired participants in the user evaluation study, highlighting their varied demographics, visual abilities, online shopping habits, and screen reader preferences.]{User Evaluation Participant Demographic Table}
\label{tab:phase2demo}
\vspace{-0.14in}
\end{table}
\vspace{-0.14in}

% \todo{add testing sequence for each participant}

% recover
\begin{table}[ht]
\centering
\begin{tabular}{|l|l|}
\hline
\rowcolor[HTML]{EFEFEF} 
\multicolumn{1}{|c|}{\cellcolor[HTML]{EFEFEF}\textbf{Testing Sequence}} & \multicolumn{1}{c|}{\cellcolor[HTML]{EFEFEF}\textbf{Participants}} \\ \hline
HTML → Tag → Original & P4, P12 \\ \hline
HTML → Original → Tag & P2, P11, P15 \\ \hline
Tag → HTML → Original & P5, P9 \\ \hline
Tag → Original → HTML & P6, P10, P14 \\ \hline
Original → HTML → Tag & P1, P7, P13 \\ \hline
Original → Tag → HTML & P3, P8 \\ \hline
\end{tabular}
\caption{\fixme{Sequence of Version Testing by Participants}}
\label{tab:sequence_participant}
\end{table}

\subsection{Study Procedure}
\label{user_study_procedure}
\fixme{We conducted a within-subjects user evaluation over Zoom, lasting 60-90 minutes. This duration included time for setting up the experiment (e.g., installing an extension), completing all three tasks across the three website versions, filling out an evaluation survey after each version (three surveys total), and engaging in think-aloud protocols and discussions with researchers to capture detailed feedback and insights.} At the beginning of the study, we obtained participants' consent for both participation and recording. We started with general questions about their previous experiences with online shopping and screen reader usage. This was followed by user testing of three different versions of the Mercari.com website: (1) the original website, (2) the LLM-regenerated HTML version implemented through Extension Option 1, and (3) the LLM-reorganized HTML tags version implemented through Extension Option 2, as described in Section~\ref{sec:phase2}.

\fixme{To prevent learning effects, we counterbalanced the sequence of versions each participant tested. The testing sequence for each participant is detailed in Table~\ref{tab:sequence_participant}. Drawing from prior research~\cite{jones2024customization}, we designed three tasks that simulate key navigation and decision-making processes, reflecting real-world shopping scenarios.

For each version, participants started by browsing the homepage to familiarize themselves with the layout of a new shopping website (Task 1). Once they completed this, they were asked to imagine they wanted to buy a blender and search for it on the site. We chose a blender as it is a gender-neutral item, helping us avoid content bias that might influence the browsing experience. Participants then reviewed the search results, exploring the products as they typically would when searching on an e-commerce website, and selected a product they were interested in exploring further (Task 2). Finally, they clicked on their chosen product and browsed the product detail page to gather the information they needed (Task 3).

During these tasks, participants were encouraged to browse as they normally would while using their screen readers and to think aloud about their experience. This included sharing how they navigated the site, any challenges they encountered, and their overall thoughts on the browsing experience.}
% To prevent learning effects, the sequence in which participants tested the different versions was randomly assigned. The testing sequence for each participant is shown in Table~\ref{tab:sequence_participant}. 
% For each version of the website, participants were first instructed to browse the homepage and explore the site according to their preferences. They were then presented with a scenario in which they were to search for and select a blender on the shopping website.

\begin{enumerate}
\item \textbf{Task 1: Homepage} Participants navigated the homepage of the shopping website using a screen reader as they normally would.

\item \textbf{Task 2: Search Page} Participants used the screen reader to search for a blender on the website and browsed through the search results.

\item \textbf{Task 3: Product Page} Participants used the screen reader to view the details of a specific product they were interested in purchasing.
\end{enumerate}

\fixme{To better understand how well each version addressed participants' accessibility needs and navigation preferences on e-commerce webpages, we designed a brief Likert-scale survey with five questions. These questions were adapted from a previous study~\cite{jones2024customization} and informed by findings from our formative study, tailored specifically to accessibility concerns and shopping scenarios.}

% After completing three tasks on each version of the webpage, participants were asked to self-report on 5-question survey to further explore the challenges and ease of use they experienced as screen reader users. They then completed a 5-question survey, with all questions rated on a 1-5 Likert scale. 

\begin{enumerate}
\item \textbf{Q1: }How would you rate your overall experience using the screen reader on this website? 
\item \textbf{Q2: }How clear and meaningful did you find the headings on this website? 
\item \textbf{Q3: }How helpful were the headings in understanding the hierarchy of content on this website? 
\item \textbf{Q4: }How efficiently could you locate key sections or information on this website using headings? 
\item \textbf{Q5: }How easy was it to access key features (e.g., search bar, product categories, account options) using your screen reader on this website?
\end{enumerate}

\fixme{Participants rated each question on a 5-point likert scale, where 1 indicated "Very Difficult / Poor" and 5 indicated "Very Easy / Good." To ensure consistent evaluation of accessibility and usability, participants completed the survey after testing each version, resulting in three sets of responses within a single interview.} After participants completed browsing all three versions, we asked follow-up questions such as how similar they found the content across the three versions, how they would compare these versions, and to rank them based on their experience. Finally, we debriefed participants about the differences between the versions and asked for additional suggestions for design improvements.

\subsection{Data Analysis}
We analyzed the user evaluation data (think-aloud and post-task interviews transcripts) using an inductive thematic analysis~\cite{braun2006using}. Two researchers independently read through each transcript, developing preliminary code set grounded in participants’ navigational preference and comparative opinions of each website version. These codes included recurring observation related to prefered heading structure, button labeling and reorganized content sequence. They met each week during the coding process to discuss about new codes and reconcile the discrepancies on code definition and scope. Ultimately, they identified approximately 50 themes, including ``Trust in Consistent Labeling'' and ``Heading Overload''.

\subsection{Quantitative Results}
% \  {I think you might need to do a Bonnferroni correction for the post-hoc multiple comparisons. some reviewers might invalidate your results if you don't correct for the family wise type I error rate} 

\subsubsection{Statistical Analysis of Likert Scale Survey Results}
\fixme{To analyze the results from the 5-question survey for each version, We conducted Wilcoxon signed-rank tests for all pairwise comparisons across five questions (Q1–Q5), comparing three different conditions: Original, HTML, and Tag. To account for the increased risk of Type I errors due to multiple comparisons, a Bonferroni correction was applied for each question, adjusting the significance threshold to \(\alpha_{\text{adjusted}} = 0.0167\). Results indicate that the HTML version consistently outperformed the Original website across all five questions, while the Tag version showed significant improvements only in the overall user experience (Q1) compared to the Original. The comparisons between HTML and Tag versions revealed a significant difference only in Q1, suggesting that HTML enhancements had a more pronounced impact on improving user experience than Tag reorganizations.}

\fixme{
\textit{Overall User Experience (\textbf{Q1}):} Participants rated the overall user experience higher for both HTML and Tag versions compared to the Original website. The Wilcoxon signed-rank tests showed significant differences between the Original website and HTML (\( W = 0 \), \( p = 0.00585 \)) and between the Original website and Tag (\( W = 0 \), \( p = 0.00585 \)). Additionally, the comparison between HTML and Tag (\( W = 0 \), \( p = 0.01170 \)) was also significant. All these p-values are below the adjusted significance threshold (\(\alpha_{\text{adjusted}} = 0.0167\)), indicating that both HTML and Tag versions significantly improved the overall navigation experience compared to the Original website.

\textit{Clarity of Headings (\textbf{Q2}):} Clarity of headings was rated higher in both HTML and Tag versions compared to the Original website. The comparison between the Original website and HTML (\( W = 0 \), \( p = 0.00585 \)) was significant, indicating improved clarity. However, the comparison between the Original website and Tag (\( W = 0 \), \( p = 0.04500 \)) and between HTML and Tag (\( W = 0 \), \( p = 0.18750 \)) were not significant after Bonferroni correction.

\textit{Understanding of Content Hierarchy (\textbf{Q3}):} Participants reported improved understanding of content hierarchy in both HTML and Tag versions compared to the Original website. The comparison between the Original website and HTML (\( W = 0 \), \( p = 0.00585 \)) was significant, indicating a better understanding. However, the comparisons between the Original website and Tag (\( W = 0 \), \( p = 0.04500 \)) and between HTML and Tag (\( W = 5 \), \( p = 0.18750 \)) were not significant after correction.

\textit{Efficiency in Locating Key Sections (\textbf{Q4}):} Option HTML outperformed the Original website in efficiency (\( W = 0 \), \( p = 0.00585 \)), which was significant. Comparisons between the Original website and Tag (\( W = 2.5 \), \( p = 0.16470 \)) and between HTML and Tag (\( W = 0 \), \( p = 0.17550 \)) were not significant after Bonferroni correction.

\textit{Ease of Access to Key Features (\textbf{Q5}):} Ease of access to key features improved in the HTML version compared to the Original website (\( W = 0 \), \( p = 0.00585 \)), which was significant. However, comparisons between the Original website and Tag (\( W = 5.5 \), \( p = 0.10260 \)) and between HTML and Tag (\( W = 0 \), \( p = 0.35100 \)) did not reach statistical significance after correction.

}

\begin{figure}[h]
    \centering
    \begin{subfigure}[t]{0.5\linewidth}
        \centering
        \includegraphics[width=\linewidth]{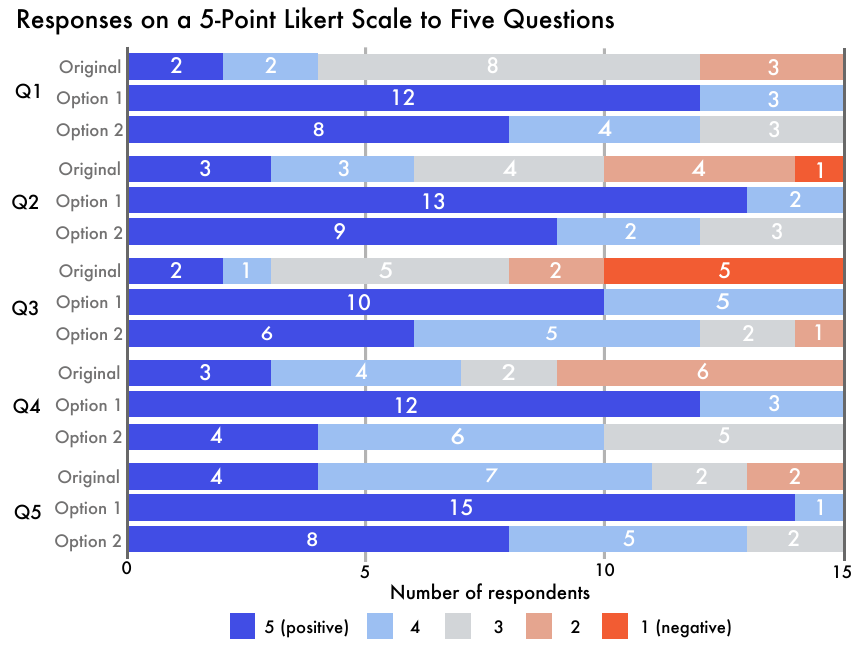}
        \caption{\fixme{The 5-point Likert scale results from Q1 to Q5, as answered by 15 participants when testing the three versions. From top to bottom, ``Original'' refers to the original website, ``Option 1'' refers to the LLM-regenerated HTML version implemented through Extension Option 1, and ``Option 2'' refers to the LLM-reorganized HTML tags version implemented through Extension Option 2.}}
        \label{fig:5likerd_result}
    \end{subfigure}
    \hfill
    \begin{subfigure}[t]{0.45\linewidth}
        \centering
        \includegraphics[width=\linewidth]{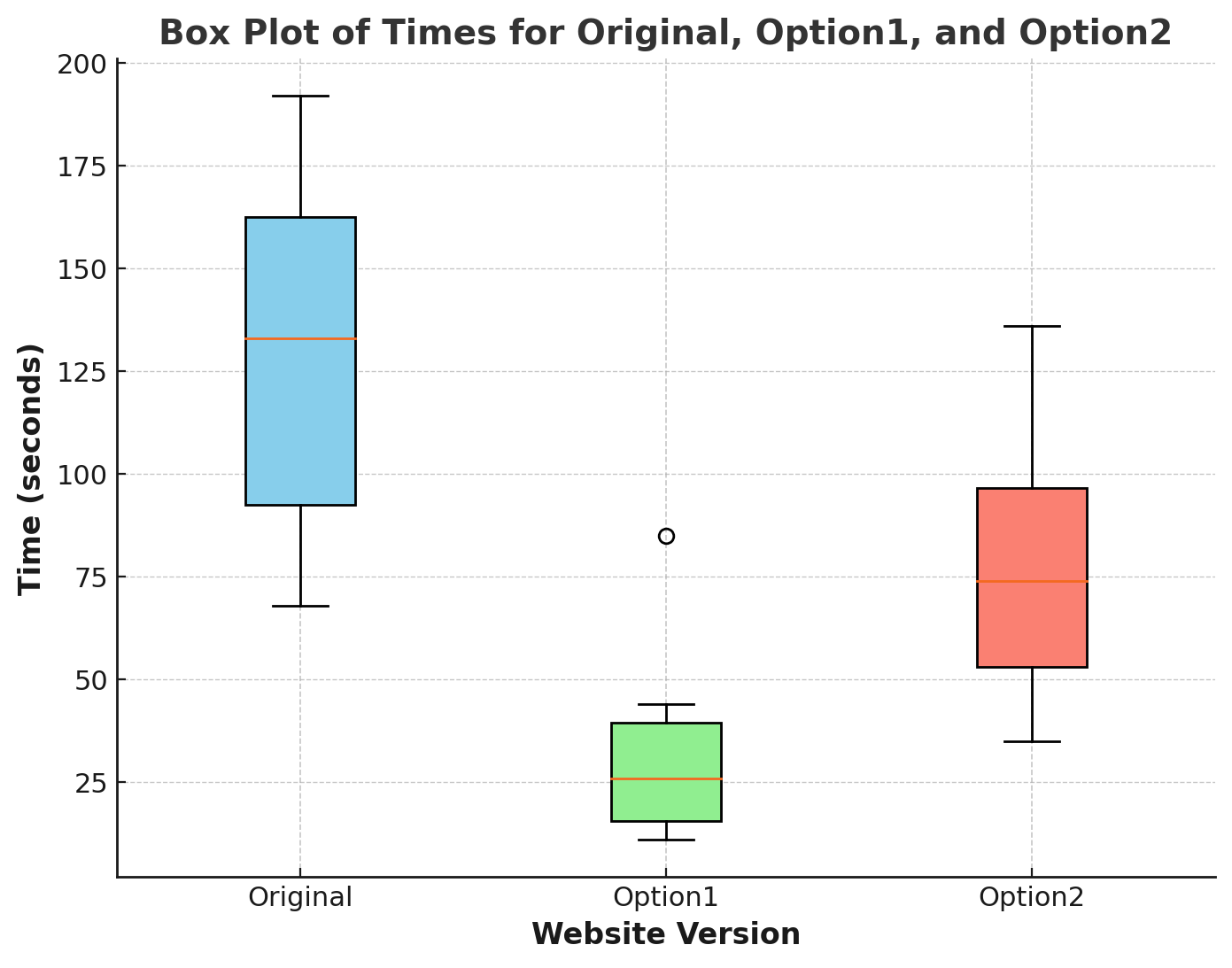}
        \caption{\fixme{Box plot comparing task completion times across the three versions (Original, Option 1, and Option 2), showing variability in performance and highlighting differences in accessibility design.}}
        \label{fig:boxplot_result}
    \end{subfigure}
    \caption{\fixme{Comparison of participant responses and task performance across the three webpage versions.}}
    \label{fig:comparison}
\end{figure}

\subsubsection{Task Completion Time in Different Version}
\fixme{We instructed participants to complete tasks 1–3 that simulated their real-life shopping experiences. They were encouraged to freely browse products without a strict goal of locating specific information or identifying a particular product on the search results page. Participants were also prompted to think aloud as much as possible during the tasks, sharing their thoughts on accessibility and navigation. Given this open-ended setup, comparing the time it took participants to complete each task (e.g., browsing the homepage or search results page) would be unreasonable, which would not accurately reflect the differences in accessibility design between versions. Some participants naturally verbalized more thoughts or explored more products on the search results page than others. However, in Task 3, while browsing the product details page, all participants consistently began by locating the product description. This behavior allowed us to calculate the time each participant spent finding the product description as a reliable quantitative metric for comparing the accessibility of different versions~\cite{islam2023spacex}. All participants naturally prioritized this step, making it a reliable and meaningful metric for comparing accessibility design across the three versions. The task completion times, as visualized in the box plot (Figure~\ref{fig:boxplot_result}), reveal significant differences across the three webpage versions: Original, Option 1 (Regenerated HTML) , and Option 2 (Reorganized HTML Tags). 

To analyze these differences, we conducted a Friedman test followed by Wilcoxon signed-rank tests for pairwise comparisons. Given that three pairwise tests were performed, we applied the Bonferroni correction to control for the increased risk of Type I errors. Specifically, the original significance level (\(\alpha = 0.05\)) was divided by the number of comparisons (\(m = 3\)), resulting in an adjusted significance level (\(\alpha_{\text{adjusted}} = 0.0167\)). The Friedman test indicated a significant overall difference among the versions (\(\chi^2(2) = 30.00\), \(p < 0.001\)). Subsequent pairwise comparisons using the Wilcoxon signed-rank test, adjusted with the Bonferroni correction, demonstrated that Option 1 significantly outperformed the Original website (\(W = 0.0\), \(p < 0.001\)) and Option 2 (\(W = 0.0\), \(p < 0.001\)). Additionally, Option 2 showed a significant improvement over the Original version (\(W = 0.0\), \(p < 0.001\)) but exhibited greater variability in task completion times. Option 1 consistently achieved the shortest task completion times, reflecting its effective structural and navigational enhancements. Option 2 provided moderate improvements but was limited by the unchanged layout, while the Original website showed the longest times and the most inconsistencies due to unstructured content and missing navigation labels. These results highlight the importance of combining structural and usability improvements to create a more accessible and efficient user experience.

% In addition to measuring the time spent locating the product description, we also tracked the error rate, which represents the number of navigation challenges participants encountered. These challenges included difficulties locating products on the search results page, missing key product details, experiencing confusion about the page layout, struggling to follow a logical flow, or needing to backtrack repeatedly to find the desired information.
}

% \subsubsection{Limitations}

\subsection{Qualitative Results}
\subsubsection{Browsing Patterns of Screen Reader Users on Unfamiliar Website}
The most commonly used navigation combination by all participants is the \textit{heading key} paired with the \textit{down arrow key}. The \textit{heading key} allows users to jump between headings on a webpage by pressing \textit{``H''}, similar to how sighted users scan visually for bold or large text to find different sections. The \textit{down arrow key} is used to navigate through HTML elements one by one, enabling users to read and explore content sequentially. When visiting a new website, participants typically began by pressing the down arrow to read line by line and discover whether the page made effective use of headings. Once they recognized headings were present, they switched to pressing \textit{heading key (H)} to move among heading levels more swiftly. If they found a heading relevant (for instance, product details labeled with an <h3> or <h4>), they returned to the down arrow key to read underlying elements in sequence. Sometimes, they may also navigate by specific heading levels (number keys 1–6). For example, if they browse a product title labeled as a heading level 2, and then want to read the product details, they might press the number 3 key to jump to the next heading level, assuming the details are structured logically under that level. In addition to the heading and down arrow keys, some participants also mentioned using the \textit{button key} and \textit{edit key} to interact with interactive elements on a webpage. The button key allows users to quickly navigate to buttons, such as \textit{``Add to Cart''} or \textit{``Submit''}, while the edit key helps them jump directly to input fields, making it easier to fill out forms or enter search queries.

\subsubsection{User Preferences for Regenerated HTML Versions}

All participants found both the regenerated HTML version and the reorganized HTML tags version to be more intuitive for navigating with a screen reader, significantly enhancing accessibility. The regenerated HTML version was especially prefered for its logical reordering of sections and enhanced navigational HTML labels. The participants' average evaluation of the screen reader browsing experience on the regenerated HTML version was 5 on a scale from 1 (Very Poor) to 5.00 (Excellent). The average rating for the reorganized HTML tags version was 4.57, while the original website received an average rating of 3.14. 13 participants reported that the regenerated HTML version provided the best experience, even though the reorganized HTML tags version also showed improvements over the original website. P6 described the navigation experience on regenerated HTML version as ``highly structured. Even for people who just started using screen readers, they can navigate quickly without confusion.'' P3 also mentioned that ``(regenerated HTML) version has a title in the heading `Purchase Options' that summarizes buttons and information related to purchasing, which is something I’ve never seen on other websites.'' One participant felt that both versions offered an equally good screen reader navigation experience, while another preferred the reorganized HTML tags version over the regenerated HTML version.

\subsubsection{Accessibility Challenges of Original Website}
While browsing the original website, participants often experienced inconvenience and confusion due to the inappropriate use of HTML tags and a layout that did not align with screen reader browsing patterns. P2 noted, \textit{``The website is cluttered, maybe like too many uses of the headings, but there's nothing I could do.''} Participants also acknowledged that this was a common problem on other mainstream shopping websites. Although it is not convenient for screen reader users, they have become accustomed to it and consider it acceptable. As P2 explained, \textit{``Amazon is an insanely cluttered website. There are so many headings. The only way to navigate Amazon is to just arrow down each item instead of using headings.''} Participants also reported frustration when important information, such as product titles, was not marked as headings. For example, P3 noted, \textit{``So the problem is the product itself is not a heading, but a lot of other information that is not necessarily needed is in headings.''} Additionally, the structure and sequence of the website posed challenges for participants in understanding and navigation. For instance, some websites place product images and reviews on the left, while product names and details are on the right. This layout may make sense from a sighted user's perspective due to visual design principles, but it creates confusion for blind or low vision users. Because review titles are also marked as headings, when users press ``H'' to navigate through headings, reviews appear before the product title. P4 expressed confusion with this structure, \textit{``It is good that they put reviews in headings, but not as the first heading we encounter, because it serves as a signal that it is already the end part of the product information.''} \fixme{P9 also expressed concern about the complexity of the layout, stating, \textit{``Whenever I hit H, it would bring me to a completely different part of the page.''}}
\vspace{-0.05in}
\subsubsection{Correct Labeling for Essential Information}

Participants perceived that both the regenerated HTML version (version 2) and the reorganized HTML tags version (version 3) improved the heading structure compared to the original website. They rated the helpfulness of headings for understanding the website's hierarchy with an average score (Q3) of 4.86 for the regenerated HTML version and 4.43 for the reorganized HTML tags version, in contrast to a score of 2.71 for the original version. Participants reported significant improvements in both versions 2 and 3 in two key aspects: (1) unnecessary headings, such as individual category names (e.g., women, men, sports), were removed. For example, P2 explained, \textit{``Those categories are just regular links [on the regenerated HTML version and reorganized HTML tags version], and it is faster to navigate through the page because you can easily skip that whole section of the website if it's not what I'm interested in.''} (2) Important information, like product names, was correctly labeled as headings. P4 mentioned, \textit{``I see that links, such as the title of the product, are also displayed as headings [on the regenerated HTML version and reorganized HTML tags version], which is quite convenient for me because now I can simply press the number key for the product.''} \fixme{P13 noted that important information appears as headings, stating,  \textit{``Headings are very much in the key part of the page, and they really help me quickly navigate, revealing that they also assist in navigation.''}}

\subsubsection{Why Regenerated HTML Works Better}
The regenerated HTML version (version 2, Q4 $\bar{x} = 4.86$) was perceived as even better than the reorganized HTML tags version (version 3, Q4 $\bar{x} = 4.28$) in terms of locating essential information and understanding the hierarchy of website content. In version 2, the LLM regenerated the entire HTML file, leading to several unique improvements that version 3 could not achieve:

\begin{enumerate}
    \item \textbf{Logical Reordering of Sections:} The regenerated HTML reorganized the page's sections into a more logical, linear sequence that better suits screen reader navigation. For example, participants observed that in the regenerated version, the reviews and seller profile appear after the product name and details, but before the purchase options. This follows a logical order that aligns with the linear browsing patterns of screen readers. In contrast, on the original website, elements were arranged for visual aesthetics—product images and reviews were positioned on the left side, which in the HTML code would be read first by screen readers, even before the product name and details. Additionally, the purchase options were presented before the product details in the HTML structure. P6 highlighted this improvement, \textit{"[The regenerated HTML version] is highly structured. Even for people who just started using screen readers, they can definitely navigate quickly without confusion."}

    \item \textbf{Addition of Summary Headings:} The regenerated HTML version added summary text to sections and aligned them as headings, which participants found extremely helpful. These changes made version 2 stand out not only from the original website but also from other shopping websites. Participants considered this a uniquely nuanced accessibility improvement. As P3 noted, \textit{``Version 2 has a title in the heading 'purchase options' that summarizes buttons and information related to purchasing, something I have never seen on any other websites. Others normally just put all buttons in headings. Now I can quickly jump through the purchase information like ``add to cart'' and ``buy now'' while browsing; that's impressive.''} \fixme{P5 also mentioned \textit{``It is very easy to read because it has both headings and lists. ''}}

    \item \textbf{Enhanced Navigation Flexibility:} The regenerated version used different labels on the same element, enriching screen readers' options for navigation. P7 mentioned that each product title is not only a heading but also a link and part of a list structure: \textit{``It's very flexible to navigate; I can use list navigation, heading navigation, or link navigation.''} \fixme{P14 noted \textit{``The number of times I had to search through headings to reach the results has significantly decreased.''}}
\end{enumerate}
\vspace{-0.14in}
\subsubsection{\fixme{User Perception on Content Similarity}}
\label{sec:user_similarity}
\fixme{When it comes to self-reported perception of content similarity, all participants noted the high similarity in content between versions, only noting the changes in design or layout of the website. For instance, P9 has mentioned that \textit{``I don't think there's any difference. It was just where they were located and how long it took to get to them.''} While P8 mentioned that \textit{``overall, it looked pretty similar.''} and similarly P10 has mentioned \textit{``It definitely seems like a lot of the same content, you know, just in different orders depending on how you click on stuff.''}}

\section{Discussion}
\label{sec:discuss}
Our formative study revealed key mismatches between shopping website designs and the navigation behaviors of screen reader users. Even when pages followed accessibility guidelines, users still faced confusion due to inconsistent HTML structure, overloaded headings, and poor labeling. These issues disrupted linear navigation, especially on unfamiliar sites, where users struggled to locate essential information within cluttered or visually prioritized layouts.

To address these barriers, we developed a GenAI-based pipeline that reorganizes and regenerates HTML to align with screen reader navigation patterns. This proof-of-concept demonstrated clear improvements in labeling, content hierarchy, and usability. In our user evaluation, participants strongly preferred the revised versions and highlighted preferences for clear section order, consistent headings, and reduced visual clutter.

These findings reinforce the need for accessibility tools that go beyond technical compliance and support the actual browsing logic of screen reader users. By grounding our system design in formative insights, we show how GenAI can deliver practical, user-centered improvements for accessible web design. We reflect on our findings to discuss limitations and opportunities for future work:

\subsection{Generative AI as a Tool for Accessibility: Supporting Developers, Not Shifting Responsibility}
The proposed Generative AI (GenAI) approach to revising HTML for better accessibility is not intended to absolve companies of their responsibility to design inclusive interfaces. While our implemented browser extension dynamically reorganizes or regenerates HTML to enhance screen reader navigation, it is not intended as a permanent solution that shifts the burden of accessibility onto users. Instead, the primary purpose of our proposed method and implementation serves as a proof-of-concept, which is to demonstrate the potential of such technology in improving web accessibility and to evaluate user preferences for future website content structures. 

Companies should take full responsibility for designing accessible websites from the outset and integrating GenAI into their development workflows. By doing so, they can ensure that accessibility improvements are implemented and controlled during the design and testing phases rather than relying on end-users to apply external tools to address these issues. As highlighted in prior research, tools like \textbf{accessibility overlays} have been criticized for creating new barriers rather than resolving existing ones, further emphasizing the need for developer-led interventions~\cite{bigham2010vizwiz, makati2024promise}. 

By leveraging GenAI responsibly within development pipelines, companies can enhance web accessibility while maintaining control over the design process. This ensures that accessibility improvements are both effective and sustainable, reducing the burden on end-users who rely on assistive technologies. For instance, future work could explore real-time feedback mechanisms that provide accessibility suggestions and quick fixes, allowing developers to continuously refine their systems and enhance their adaptability across different browsing environments. Specifically, the pipeline could be used to generate or reorganize a developer's HTML product and visualize the changes and reasoning in a sidebar, enabling developers to make informed decisions. This system could then be further evaluated and improved based on developer feedback. By extending the pipeline’s capabilities in these ways, the tool could become both a practical solution for developers looking to create more inclusive web experiences.

\subsection{Broadening GenAI Accessibility Enhancements to Diverse Websites}
Generative AI (GenAI) offers significant potential for improving web accessibility across diverse contexts, extending beyond the e-commerce platforms explored in this study. While our research primarily focuses on shopping websites, the challenges identified such as poor information hierarchy, inadequate labeling, and navigation inefficiencies are prevalent across many types of websites. These barriers can impact screen reader users navigating other contexts, including news platforms, educational resources, and service-based websites. For example, news platforms often contain complex layouts with articles, headlines, sidebars, and advertisements competing for attention. GenAI could reorganize these sections to create a logical flow that enhances readability for screen reader users. Prior research emphasize the importance of clear and concise content presentation for visually impaired users navigating dense information spaces, which aligns with the potential application of GenAI in this domain~\cite{stangl2021going, wang2022makes}.

However, applying GenAI across different contexts is not without challenges. Websites with dynamic or user-generated content may require real-time adjustments to maintain accessibility standards. Additionally, different website types have distinct user requirements; for instance, news platforms prioritize readability while service-based sites focus on task efficiency. Ensuring compliance with WCAG guidelines while adapting to diverse contexts is also critical for widespread adoption of GenAI solutions. Future research should explore the generalizability of GenAI-powered tools by conducting studies on various website categories to identify specific barriers and refine GenAI models accordingly. Hybrid solutions combining automated GenAI outputs with manual adjustments could address context-specific needs while maintaining flexibility in accessibility improvements. Furthermore, developing tools that allow developers to customize GenAI-generated outputs based on user feedback and accessibility requirements would enhance the practical application of these technologies.
\section{Conclusion}

Accessibility in online shopping remains an open problem for blind or low vision screen reader users. In this study, we investigated the issues experienced by screen reader users when shopping online and developed a web browser plugin that applies GenAI to improve the accessibility of the website for screen readers. Our evaluation shows improvement (as reported by accessibility checkers and self-reported by users) in the navigation and labeling of items and can potentially be used by both users and developers to improve the accessibility of their online shopping experience. 

We believe that our work proves that application of GenAI and LLMs in particular can automatically improve the web browsing accessibility and reduce the burden on both users and developers. We believe that our work contributes to the topic of web development and accessibility in the coming age of ubiquitous human-AI interactions, and we hope that our study would contribute to the work of researchers and practitioners in the fields of AI and HCI.

\bibliographystyle{ACM-Reference-Format}
\bibliography{reference}
% \usepackage[backend=biber,style=ACM-Reference-Format]{biblatex}
%\bibliography{sample-base}

\newpage

\appendix

\appendix
\section{Appendix}
\appendix

\section{Implementation Details}

\subsection{System Configuration and LLM Setup}
\label{sec:llm_config}

The Chrome extension's background script uses the OpenAI API to access \texttt{GPT-4o}~\cite{gpt4o}, implemented through the browser's \texttt{fetch} Web API~\cite{openai_jsonmode}. We selected \texttt{GPT-4o} for both regeneration (Option 1) and tag reorganization (Option 2) due to its large input (128,000 tokens) and output (16,384 tokens) capacities~\cite{gpt4o-tokens}, which allow extensive content processing while maintaining structure and context.

To optimize for deterministic and consistent outputs, the following parameters were used: \texttt{temperature = 0.2}, \texttt{max\_tokens = 16,384}, \texttt{top\_p = 1}, \texttt{frequency\_penalty = 0}, and \texttt{presence\_penalty = 0}. A low temperature helps maintain formatting consistency in HTML, while top-p = 1 ensures complete content generation. The penalties are set to 0 to avoid suppressing necessary recurring elements, such as repeated headings or button labels.

Each webpage required 1–5 minutes to process, with around 220,000 tokens consumed per session (combined input and output). According to OpenAI’s pricing\footnote{\url{https://openai.com/api/pricing}}, the cost per page ranged from \$0.50 to \$2.20. This could be reduced in production by caching outputs or using lighter models.

\subsection{Option 1: Regenerated HTML Version}
\label{sec:option1_workflow}

\subsubsection{Step 1: HTML Collection}

We used a Google Chrome extension to extract complete webpage HTML. A \textbf{content script} interacts directly with the Document Object Model (DOM) and retrieves the full HTML using \\ \texttt{document.documentElement.outerHTML}. The captured HTML, including dynamic JavaScript-rendered elements, is passed to the \textbf{background script}.

\subsubsection{Step 2: Preprocessing}

The background script segments the raw HTML based on DOM structure to ensure each chunk fits within the token limit of \texttt{GPT-4o}. Segmentation is done at logical HTML boundaries (e.g., top-level \texttt{<div>} or \texttt{<section>} blocks) to preserve context. Tokenization ensures each chunk respects model constraints.

\subsubsection{Step 3: Sequential LLM Calls}

Each HTML chunk is processed using OpenAI’s \texttt{chat/completions} API. A system prompt, user prompt, and assistant context (see Section~\ref{sec:prompts}) guide the model to generate an accessible HTML version. The assistant message includes prior output to maintain continuity across chunks.

\subsubsection{Step 4: Similarity Check}

Before injecting the generated output into the live webpage, our system conducts a similarity check to ensure that the transformed HTML retains the core content and meaning of the original page. To maintain consistency between implementation and evaluation, we used the same metric—\textbf{Aggregated Semantic Similarity}—as described in Section~\ref{sec:evaluation}.

We compute semantic similarity by extracting all screen-reader-accessible HTML elements (e.g., visible text, ARIA labels, alt attributes) from both the original and generated versions, concatenating them, and comparing their embeddings using cosine similarity. The transformation is accepted only if the aggregated semantic similarity score meets or exceeds a \textbf{90\% threshold}. This threshold was chosen based on preliminary testing to ensure the balance between allowing accessibility-enhancing changes (e.g., restructuring or paraphrasing) and preserving the original meaning. If the score falls below the threshold, the system automatically reprocesses the content to improve alignment.

\subsubsection{Step 5: Page Replacement}

Once the final HTML passes the checks, it replaces the original webpage DOM using JavaScript injection. All functional elements (e.g., forms, buttons, links) remain intact. If any issues are detected post-rendering, additional regeneration is triggered until the results are usable and accessible.

\subsection{Option 2: Reorganized HTML Tags}
\label{sec:option2_workflow}

\subsubsection{Step 1: DOM Extraction and JSON Serialization}

Option 2 focuses on minimally modifying the webpage structure while improving accessibility. HTML is extracted similarly via the extension, but converted into a structured JSON format for fine-grained tag-level manipulation. This includes tag names, attributes, and content.

\subsubsection{Step 2: Preprocessing}

We filter out unnecessary elements (e.g., styles, scripts) and chunk the JSON for compatibility with the \texttt{GPT-4o} model’s 32,768-token input window. Segments are created at logical DOM subtrees to preserve structural coherence.

\subsubsection{Step 3: LLM-Powered Tag Optimization}

Each chunk of JSON is sent via the same Web API using system and user prompts (Section~\ref{sec:prompts}) tailored to instruct the model to adjust headings, labels, and ARIA attributes for improved screen reader navigation—without introducing or removing any content.

\subsubsection{Step 4: Similarity Check}

Although Option 2 only modifies the HTML tags and not the visible content or layout, we still perform a similarity check to ensure that the tag reorganization does not unintentionally alter or omit content relevant to screen reader users. We apply the same \textbf{Aggregated Semantic Similarity} metric used in Option 1 and in our evaluation (Section~\ref{sec:evaluation}) to compare the original and tag-updated versions.

We extract all accessible HTML elements (including headings, labels, and text content) from both versions, concatenate them, and compute cosine similarity between their semantic embeddings. The updated version is accepted only if the aggregated semantic similarity score is at least \textbf{90\%}. This threshold ensures that any tag changes preserve the original content's meaning and utility for screen reader users. If the similarity score is below the threshold, the chunk is flagged for reprocessing.

\subsubsection{Step 5: Live DOM Update}

Once validated, the extension updates the active page using a \texttt{replaceTags()} function, replacing only the tags while keeping the text and visual layout unchanged. All interactive elements remain fully functional, now better marked up for screen reader interpretation.

\subsection{Prompt Design}
\label{sec:prompts}

In both options, LLM prompts are composed of a \textbf{System Prompt} to establish behavior, a \textbf{User Prompt} to pass specific HTML or JSON input, and an \textbf{Assistant Message} to maintain continuity across chunks.

\subsubsection{Prompts for Option 1 (Regeneration)}

\textbf{System Prompt:}
\begin{quote}
Your task is to enhance the accessibility of the shopping website for blind screen reader users. Screen reader users navigate websites sequentially and use shortcuts to jump across headings and links. Remove non-essential elements (e.g., pictures, <style>, <script>) and ensure all content from the original HTML is included and properly structured.
\end{quote}

\textbf{User Prompt:}
\begin{quote}
Generate executable, text-only HTML optimized for screen reader users. Reorganize headings to reflect information hierarchy and enhance clarity. Prioritize content that is most relevant for screen reader navigation.
\end{quote}

\textbf{Assistant Prompt:}
\begin{quote}
This is the previous response from the system: \textit{<HTML content>}
\end{quote}

\textbf{Follow-up User Prompt:}
\begin{quote}
This is part X of the HTML document. Please continue processing it while maintaining the previous structure. \textit{<next HTML segment>}
\end{quote}

\subsubsection{Prompts for Option 2 (Reorganization)}

\textbf{System Prompt:}
\begin{quote}
You are an accessibility expert. Screen reader users navigate content sequentially, relying on well-structured headings, landmarks, and clear labels. Your task is to revise the tags of an existing webpage to improve accessibility without modifying its visual layout or content.
\end{quote}

\textbf{User Prompt:}
\begin{quote}
Please adjust tags to enhance accessibility for screen reader users. Do not introduce or remove elements. Only reorganize or relabel existing tags based on structural clarity and semantic accuracy.
\end{quote}

\textbf{Assistant Prompt:}
\begin{quote}
This is the previous response from the system: \textit{<JSON content>}
\end{quote}

\textbf{Final Message (if needed):}
\begin{quote}
This is the remaining JSON content that needs to be processed. Ensure completeness and do not duplicate any previously generated elements.
\end{quote}

\subsection{Example Accessibility Enhancements}
\label{appendix:examples}
To demonstrate the types of accessibility enhancements applied by our system, we summarize typical modifications made under each version. 
\subsubsection{Option 1: Regenerated HTML}
\begin{itemize}
    \item Consistent and Semantic Heading Structure: The regenerated HTML ensures a clear and consistent use of heading tags (<h1>, <h2>, <h3>, etc.) throughout the pages, following a logical order. This makes it easier for screen reader users to understand the page layout and navigate to relevant sections. For example, on the homepage, the <h1> tag is appropriately used for the main title, ``Shop Popular Items,'' while sub-sections such as ``Featured Categories'' and ``Today's Picks'' use <h2>. This prevents the confusion caused by random or missing headings in the original version.
    \item Replace headings to distinguish and highlight information: The content is logically organized with unnecessary elements removed from headings, and summarizing titles added to enhance clarity and reduce clutter. This makes it easier for users to focus on the relevant information without distractions. For example, on the search result page, instead of showing all filter options and categories as separate headings, the LLM-regenerated version groups them under a single ``Filter Results'' section, allowing users to navigate efficiently.
    \item Reordered Sections for Logical Screen Reader Browsing Sequence: The order of sections on each page is rearranged to provide a more logical flow for screen reader users. This adjustment ensures that the most important information is encountered first, reducing the need for excessive navigation. For example, on the product details page, the regenerated HTML places ``Product Name'' and ``Product Details'' before sections such as ``Reviews'' and ``Similar Products.'' This contrasts with the original layout where reviews could appear before key product information. This change aligns better with the linear browsing patterns of screen reader users, who prefer accessing critical information first.
    \item Enhanced Use of ARIA Roles and Attributes: ARIA roles (e.g., role=``navigation'', role = ``main'', role = ``banner'') and attributes (e.g., aria-label, aria-labelledby) are added to provide additional context for assistive technologies. This allows screen readers to better interpret the purpose of various sections and elements. For exmaple, on the product details page, buttons such as ``Add to Cart'' are given specific aria-label attributes (aria-label = ``Add item to cart''), enhancing clarity for screen reader users who cannot see the button's text.
\end{itemize}

\subsubsection{Option 2: Reorganized HTML Tags Version
}
\begin{itemize}
    \item Replace headings to distinguish and highlight information: this version focuses on ensuring that headings are used appropriately to reflect the content hierarchy, removing unnecessary elements from being tagged as headings and adding key information into headings. Uses appropriate headings (<h1>, <h2>, <h3>, etc.) and paragraphs (<p>) to present content hierarchically and semantically. For example, the product name ``oster blender'' is an <h1>, ensuring it stands out as the primary focus. 
    \item Enhanced Heading Hierarchy for Logical Screen Reader Navigation: This version focuses on adjusting the levels of headings to emphasize the content's structure and hierarchy without altering the page layout. By refining the heading levels, it provides a clearer and more logical flow for screen readers. For example, on the search result page, the filter sections such as ``Filter by,'' ``Status,'' ``Item origin,'' and ``Size'' are clearly marked with <h2>, <h3>, and <h4> tags, providing a hierarchical structure that is more accessible for screen readers and improves the overall user experience.
\end{itemize}\label{appendix}

%would be nice to put a table here

% \begin{acks}

% \end{acks}


\end{document}